\documentclass[12pt]{iopart}
\usepackage{graphicx}
\usepackage{iopams}

\newcommand{\Zc}{{\cal Z}}
\newcommand{\xv}{\mbox{\boldmath$x$}}

\begin{document} 
\title{Symmetry, complexity and multicritical point of 
the two-dimensional spin glass}
\author{
Jean-Marie Maillard\dag, Koji Nemoto\ddag  \, and Hidetoshi Nishimori\S}

\address{\dag LPTHE, Tour 24, 5 \`eme \'etage, case 7109, 2 Place Jussieu,
75251 Paris Cedex 05, France}

\address{\ddag Division of Physics, Hokkaido University,
 Sapporo 060-0810, Japan}

\address{\S Department of Physics, Tokyo Institute 
of Technology, Oh-okayama,
Meguro-ku, Tokyo 152-8551, Japan}

\begin{abstract}
We analyze models of spin glasses on the two-dimensional square lattice by 
exploiting symmetry arguments.
The replicated partition functions of the Ising and related spin glasses
are shown to have many remarkable symmetry properties 
as functions of the edge Boltzmann
factors.  It is shown that the applications of
 homogeneous and Hadamard inverses
to the edge Boltzmann matrix indicate reduced complexities when the
elements of the matrix satisfy certain conditions, 
suggesting that the system
has special simplicities under such conditions.
Using these duality and symmetry arguments
we present a conjecture on the exact location 
of the multicritical point
in the phase diagram.
\end{abstract}

\section{Introduction}
Properties of spin glasses are well understood in mean-field models,
which show such anomalous behaviour as replica symmetry breaking,
multivalley structure, and slow dynamics \cite{Mezard,Young}.
The difficult problem of whether or not these mean-field predictions apply
to realistic finite-dimensional systems is still largely unsolved,
and active investigations are carried out mainly
using numerical methods \cite{Young}.
Very few systematic analytical work exists, an exception being a symmetry
argument using gauge invariance to derive the exact internal energy and
several other exact/rigorous relations \cite{Nishimori81,Nishimori01}.

In the present paper we develop another type of symmetry argument
for models of spin glasses in two dimensions under the replica formalism.
Our analyses reveal a variety of invariance properties 
of the replicated partition
function under transformations of the edge Boltzmann factors, a notable example of which
is the duality transformation.

Also discussed are complexities of the edge Boltzmann 
matrix under inversions.
It is well established that integrable systems such as 
the standard scalar Potts model
have remarkably reduced complexities of the edge Boltzmann matrix when the
parameters satisfy integrability conditions \cite{dAuriac}.
Analogous (but not the same) behaviour is observed in our present problem,
suggesting simplified (if not integrable) properties of the systems in 
restricted regions in the phase diagram.

The results concerning symmetry properties are used to present
 a conjecture on the location
of the multicritical point in the phase diagram.  Although 
the argument is not a mathematically 
complete proof, we expect the
prediction to be exact for several reasons including agreement 
with numerical results
in many different models and satisfaction of inequalities.

It should be remembered that most of the analyses are for
 real replicas, that is,
the number of replicas is a positive integer.  The quenched 
limit of vanishing number of
replicas is discussed in relation only to limited cases including the conjectured location of the
multicritical point.

In the next section symmetries of the system are derived. 
The replicated partition function is demonstrated to be invariant under several
different types of transformations of the edge Boltzmann factors.
It is shown that a special subvariety exists which satisfies
remarkable enhanced symmetries.
Complexities under applications
of matrix inverses are treated in section \ref{sec:complexity}.
Here also the same subvariety as above is seen to have reduced complexities,
suggesting its special role.
A conjecture on the multicritical point is presented 
in section \ref{sec:MCP} and numerical evidence supporting the conjecture is
discussed.
The final section is devoted to summary and discussions.

\section{Symmetries of the partition function}

In this section we first derive the expression of the 
replicated partition function
in terms of the edge Boltzmann factors.
Then the symmetries and related properties of the 
partition function are discussed.
The arguments are first developed for the $\pm J$ Ising model and then are
generalized to a broader class of models 
(such as the Gaussian spin glass and random chiral Potts model)
in the last part of this section.
\subsection{Replicated partition function}
Let us start by considering the Hamiltonian of the $\pm J$ Ising model
 \begin{equation}
   H=-\sum_{\langle ij \rangle}J_{ij} S_i S_j,
    \label{Hamiltonian_Ising}
  \end{equation}
where $J_{ij}$ is ferromagnetic $J (>0)$ with probability $p$
 and antiferromagnetic $-J$ with $1-p$.
The sum extends over nearest neighbours on the square lattice.
The randomness in  $J_{ij}$ will be treated by the replica method.
We will mainly consider the case of positive integer $n$, the number of replicas.
Some results for the quenched limit $n\to 0$ will be discussed 
in subsequent sections.
The constraints of Ising spins and the $\pm J$ distribution 
of the interactions
will be relaxed later in the present section.

After the average over bond configurations, the system 
becomes spatially homogeneous\footnote{
except for boundary effects which are irrelevant to thermodynamic properties
and will be ignored
}, and
the partition function
\begin{equation}
     [Z^n]_{\rm av} \equiv Z_n,
     \label{Zn_def}
\end{equation}
where the square brackets with subscript `av' denote the
 configurational average, is specified uniquely
by the edge Boltzmann matrix representing neighbouring interactions.
For example, when $n=1$ (the annealed model), the edge Boltzmann matrix is
\begin{eqnarray}
 p \cdot \left[ 
   \begin {array}{cc} 
      w_0&w_1\\
       \noalign{\medskip}w_1&w_0
   \end {array} 
   \right] +
  (1-p) \cdot \left[ 
   \begin {array}{cc} 
      w_1&w_0\\
       \noalign{\medskip}w_0&w_1
   \end {array} 
   \right] 
   \equiv
   \left[ 
   \begin {array}{cc} 
      x_0&x_1\\
       \noalign{\medskip}x_1&x_0
   \end {array} 
   \right] \equiv A_1,
   \label{BF1}
\end{eqnarray}
where $w_0=e^{K}, w_1=e^{-K}$ with $K=J/k_B T$, and $x_0$ 
denotes the edge Boltzmann factor
for parallel configuration of neighbouring spins whereas 
$x_1$ is for antiparallel spins.
Similarly, for $n=2$  we have the edge Boltzmann matrix
\begin{eqnarray}
 p \cdot \left[ 
   \begin {array}{cc} 
      w_0&w_1\\
       \noalign{\medskip}w_1&w_0
   \end {array} 
   \right] 
   \otimes 
   \left[ 
   \begin {array}{cc} 
     w_0&w_1\\
     \noalign{\medskip}w_1&w_0
   \end {array}
    \right] \, + \, \,(1-p) \cdot  
   \left[
     \begin {array}{cc} 
          w_1&w_0\\
         \noalign{\medskip}w_0&w_1
     \end {array} \right] \otimes 
   \left[
    \begin {array}{cc} 
      w_1&w_0\\
      \noalign{\medskip}w_0&w_1
    \end {array} \right].
    \label{BF2}
\end{eqnarray}
This is a $\,  4 \times 4$ matrix of hierarchical structure
\begin{eqnarray}
   \left[ 
   \begin {array}{cccc} 
        x_0&x_1&x_1&x_2\\
        \noalign{\medskip}x_1&x_0&x_2&x_1\\
        \noalign{\medskip}x_1&x_2&x_0&x_1\\
        \noalign{\medskip}x_2&x_1&x_1&x_0
    \end {array} \right] \equiv A_2 
    =
    \left[
    \begin {array}{cc} 
      A_1&B_1\\
      \noalign{\medskip}B_1&A_1
    \end {array} \right]
    \label{BF3}
\end{eqnarray}
with $x_0=pw_0^2+(1-p)w_1^2,\, x_1=pw_0w_1 +(1-p)w_1 w_0$
 and $x_2=pw_1^2+(1-p)w_0^2$.
The $\,  2 \times 2$ matrix
$B_1$ is obtained from $A_1$ by replacing $x_0$ and $x_1$ with $x_1$ and $x_2$, respectively:
$B_1=A_1(x_0\to x_1, x_1\to x_2)$.
The element $x_0$ represents the two parallel spin pairs on neighbouring sites
(such as $++$ for replica 1 and  $++$ for replica 2),
$x_1$ is for one parallel and one antiparallel pairs (like $++, +-$),
and $x_2$ corresponds to two antiparallel pairs
($+-, -+$ for example).

The edge Boltzmann matrix of larger $n$ can be derived by recursions.
The general formula is 

\begin{eqnarray}
   \label{general_BM}
 p \cdot \left[ 
    \begin {array}{cc} 
         w_0&w_1\\
      \noalign{\medskip}w_1&w_0
    \end {array} 
   \right] ^{\otimes n} 
   + \, \,(1-p) \cdot  \left[
      \begin {array}{cc} 
         w_1&w_0\\
      \noalign{\medskip}w_0&w_1
    \end {array} 
   \right] ^{\otimes n} 
   \equiv A_n =
   \left[ 
    \begin {array}{cc} 
         A_{n-1}&B_{n-1}\\
      \noalign{\medskip}B_{n-1}&A_{n-1}
    \end {array} 
   \right],
\end{eqnarray}
where $B_{n-1}=A_{n-1}(x_k\to x_{k+1}; k=0,1,\cdots ,n-1)$.
The partition function (\ref{Zn_def})  is a function of the matrix elements of $A_n$,
 \begin{equation}
  Z_n(x_0, x_1,\cdots , x_n).
 \end{equation}
Here $x_k$ denotes the Boltzmann factor of the spin configuration
with $k$ antiparallel spin pairs among $n$ neighbouring pairs.

In the symmetry arguments developed subsequently, we will often consider
the case where the elements $x_k$ are independent of each other although
they are originally related through the parameters of the $\pm J$ 
Ising model ($K$ and $p$)
by the relations
 \begin{equation}
  \begin{array}{rcl}
  \displaystyle
   x_0 &=& p w_0^n +(1-p) w_1^n 
      \label{generalx0}\\
   x_1 &=& p w_0^{n-1} w_1 +(1-p) w_1^{n-1} w_0
       \label{generalx1}\\
   x_2 &=& p w_0^{n-2} w_1^2 +(1-p) w_1^{n-2} w_0^2
      \label{generalx2} \\
  && \vdots \\
   x_n &=& p w_1^{n} +(1-p) w_0^{n}.
   \end{array}
   \label{generalxn}
 \end{equation}
An advantage to consider generic points in the space spanned by the
elements of the edge Boltzmann matrix
\begin{equation}
{\cal S}=\{ \xv =(x_0,x_1,x_2,\cdots , x_n)|x_k\in \mathbb{R}~(k=0,1,\cdots ,n)\}
\end{equation}
is that we can discuss various well-known
models such as the $2^n$-state standard scalar Potts model which has
$x_1=x_2=x_3=\cdots =x_n$:
All but one ($x_0$) spin configurations have the same Boltzmann factors.
The $n$-replicated $\pm J$ Ising model with the edge Boltzmann factor
(\ref{generalxn}) lies on a two-dimensional submanifold $\cal T$ of $\cal S$:
\begin{equation}
 {\cal T}=\{ \xv \in {\cal S}| x_k=pw_0^{n-k}w_1^k+(1-p)w_1^{n-k}w_0^k~(k=0,1,\cdots ,n)\}
\end{equation}
\subsection{Duality}
Some spin systems in two dimensions have invariance
 properties under duality transformations.
The formulation of duality by Wu and Wang \cite{Wu-Wang}
 is particularly useful for our
problem having the edge Boltzmann matrix (\ref{general_BM}) of hierarchical structure.
According to these authors, the dual Boltzmann factors are derived simply by
Fourier sums applied to each $\,  2 \times 2$ block
(corresponding to each replica) of the edge Boltzmann matrix
(\ref{general_BM})\footnote
{
Fourier sums for the two-component (Ising) case are just the sum and difference of two elements.
}.
The simplest case is the annealed model $n=1$ of equation (\ref{BF1}):
Its dual Boltzmann factors are the sum and difference of the original Boltzmann factors
with appropriate normalization,
 \begin{equation}
  \begin{array}{rcl}
    \sqrt{2}\, x_0^*&=& x_0+x_1 \\
    \sqrt{2}\, x_1^*&=& x_0-x_1 .
  \end{array}
  \label{dual_1}
 \end{equation}
As an example, when the system is purely ferromagnetic $p=1$ in equation (\ref{BF1}),
if we define the dual coupling $K^*$ by $e^{-2K^*}=x_1^*/x_0^*$
(remembering $e^{-2K}=x_1/x_0=w_1/w_0$), we have from equation
(\ref{dual_1}) the familiar duality relation of the ferromagnetic Ising model,
  \begin{equation}
    e^{-2K^*}=\frac{w_0-w_1}{w_0+w_1}=\tanh K.
  \end{equation}
In the case of $n=2$ with equations (\ref{BF2}) and (\ref{BF3}), it is necessary
to generate combinations of sums and differences of appropriate matrix elements
to obtain the dual Boltzmann factors,
\begin{equation}
 \begin{array}{rcl}
   2\, x_0^* &=&(x_0+x_1)+(x_1+x_2) =x_0+2x_1+x_2 \\
   2\, x_1^* &=&(x_0-x_1)+(x_1-x_2) =x_0-x_2 \\
   2\, x_2^* &=&(x_0-x_1)-(x_1-x_2) =x_0-2x_1+x_2 .
 \end{array}
 \label{dual_2}
\end{equation}

The formula for general $n$ with the edge Boltzmann matrix (\ref{general_BM}) is
\begin{eqnarray}
 \Delta_n\, = \, \, \, \, 
  {{1} \over {2^n}} 
  \left[ 
       \begin {array}{cc}
          1&1\\
          \noalign{\medskip}1&-1
       \end {array}
  \right]^{\otimes n} 
              \cdot
 \left[ 
        \begin {array}{cc} 
            A_{n-1}&B_{n-1}\\
            \noalign{\medskip}B_{n-1}&A_{n-1}
         \end {array} \right]
    \cdot
 \left[ 
      \begin {array}{cc}
            1&1\\
            \noalign{\medskip}1&-1
       \end {array}
       \right]^{\otimes n},
       \label{general_diagonal}
\end{eqnarray}
where the entries of the diagonal matrix  $\, \Delta_n\, $
are the dual Boltzmann factors $2^{n/2} x_m^{*}\, $ in a certain order.
More explicitly, the dual Boltzmann factors, which are linear combinations of the
original Boltzmann factors, can be written from equation (\ref{general_diagonal}) as
\begin{equation}
  2^{n/2} x_m^{*} \, = \, \, \, \sum_{k=0}^n  D_m^{k} \,   x_k, 
  \label{x_star}
\end{equation}
where the $\, D_m^{k}$ are the coefficients of the expansion
of  $\, (1-t)^m \cdot (1+t)^{n-m}\, $ :
\begin{equation}
 (1-t)^m \cdot (1+t)^{n-m} \, = \, \, 
\sum_{k=0}^n   D_m^{k}  \cdot  t^k,
 \label{t-expansion}
\end{equation}
that is,
\begin{equation}
 D_m^k = \sum_{l=0}^k (-1)^l {m \choose l}{n-m \choose k-l}.
 \label{Dmk}
\end{equation}
To understand equation (\ref{t-expansion}) we first note that $m$ antiparallel pairs are chosen
to generate $x_m^*$ for dual spin pairs, which is reflected in the parameter $m$ on
the left-hand side of this equation.
This is equivalent to choosing $m$ of the second rows $(1, -1)$ from the direct product
of $n$ $\,  2 \times 2$ matrices in equation (\ref{general_diagonal})\footnote
{
As seen in equation (\ref{dual_1}), an antiparallel spin pair in the dual space
corresponds to the difference (the row $(1, -1)$) of two original Boltzmann factors.
}.
Then we choose $k$ antiparallel pairs of original spins as in equation (\ref{Dmk})
to obtain the coefficient of $x_k$.

On the square lattice the partition function remains invariant under the duality
transformation of edge Boltzmann factors \cite{Wu-Wang}:
 \begin{equation}
  Z_n(x_0, x_1, x_2, \cdots ,x_n)=Z_n(x_0^*, x_1^*, x_2^*, \cdots ,x_n^*)
  ~~~(\xv \in {\cal S}),
  \label{Zn_duality}
\end{equation}
apart from a trivial factor $2^n$ and boundary effects, both of which are
irrelevant to thermodynamic properties and will be ignored in this paper.
Note that the symmetry under duality (\ref{Zn_duality}) is valid for any values of
the edge Boltzmann factors $x_0, x_1, x_2, \cdots ,x_n$, which do not necessarily
satisfy the relation (\ref{generalxn}) of the $\pm J$ Ising model
as indicated by the symbol $\xv \in {\cal S}$ at the end of equation (\ref{Zn_duality}).

It will be useful to write explicitly the general duality transformation (\ref{x_star})
applied to the specific case of the $\pm J$ Ising model for later use,
namely, for the case $\xv \in {\cal T} (\subset {\cal S})$.
Again we first write the formula for the simple case of $n=2$ 
written in equation (\ref{dual_2}) so that the reader understands the structure:
 \begin{equation}
  \begin{array}{rcl}
   \displaystyle
   2x_0^{*}&=& p(w_0+w_1)^2 +(1-p)(w_1+w_0)^2=(w_0+w_1)^2 \\
   2x_1^{*}&=& p(w_0+w_1)(w_0-w_1) +(1-p)(w_1+w_0)(w_1-w_0)\\
   &= &(2p-1)(w_0+w_1)(w_0-w_1) \\
   2x_2^{*}&=& p(w_0-w_1)^2 +(1-p)(w_1-w_0)^2=(w_0-w_1)^2.
   \end{array}
   \label{xms}
 \end{equation}
The general formula is
  \begin{equation}
   \begin{array}{rcl}
   \displaystyle
   2^{n/2} x_{2m}^{*} &=& (w_0+w_1)^{n-2m}(w_0-w_1)^{2m} \\
   2^{n/2} x_{2m+1}^{*} &=&(2p-1)(w_0+w_1)^{n-2m-1}(w_0-w_1)^{2m+1}. \\
  \end{array}
  \label{xmstar_pmJ}
\end{equation}
%
\subsection{Symmetries under sign changes of coupling constants}
In the present subsection the edge Boltzmann factors are regarded as functions
of the parameters of the $\pm J$ Ising model following equation (\ref{generalxn}), $\xv \in {\cal T}$.
The partition function $Z_n$ is invariant under the change of the sign of
coupling constant $K\to -K$ at all bonds if the system is on the square lattice.
This change of the sign exchanges $w_0$ and $w_1$, and consequently,
according to equation (\ref{generalxn}), $x_k$ is exchanged with $x_{n-k}$:
 \begin{equation}
  Z_n(x_0, x_1, x_2, \cdots ,x_n)= Z_n(x_n, x_{n-1}, x_{n-2}, \cdots ,x_0)~~~(\xv \in {\cal T}).
  \label{Zn_inv1}
 \end{equation}

Combination of two symmetries, duality (\ref{Zn_duality}) and change of sign of $K$
(\ref{Zn_inv1}), leads to another symmetry.  On the right-hand side of equation
(\ref{t-expansion}) the change of sign of $t$ is equivalent to the change
of the sign of $D_m^k$ for odd $k$.
The left-hand side of equation (\ref{t-expansion}) suggests that
the change of sign of $t$ is also realized by exchange of $m$ and $n-m$.
Then we conclude in equation (\ref{x_star}) that the change of sign
of $x_k$ for odd $k$ on the right-hand side
 \begin{equation}
  D_m^k \, x_k \to (-1)^k D_m^k \, x_k
 \end{equation}
should be performed simultaneously with the exchange of $x_m^*$ and $x_{n-m}^*$ to
keep this duality equation valid.
Using duality (\ref{Zn_duality}) we therefore find the following symmetry,
 \begin{eqnarray}
  & & Z_n(x_0, -x_1, x_2, -x_3,\cdots , (-1)^n x_n) \nonumber \\
  &=& Z_n(x_n^*, x_{n-1}^*,\cdots ,x_1^*, x_0^*)  \nonumber \\
  &=& Z_n(x_0^*, x_1^*, \cdots ,x_{n-1}^*, x_n^*) \nonumber\\
  &=& Z_n(x_0, x_1, \cdots , x_n)~~~(\xv \in {\cal T}).
 \end{eqnarray}
The second equality comes from equation (\ref{Zn_inv1}) and the final
relation is duality (\ref{Zn_duality}).

When the number of replicas is even $n=2q$, the partition function has additional symmetry
which exchanges $x_k$ with  $x_{2q-k}$ for $k$ odd only:
 \begin{eqnarray}
  &&Z_{2q}(x_0, x_1, x_2, x_3, x_4, \cdots ,x_{2q-3}, x_{2q-2}, x_{2q-1}, x_{2q})
  \nonumber\\
  &=&
  Z_{2q}(x_0, x_{2q-1}, x_2, x_{2q-3}, x_4, \cdots ,x_3, x_{2q-2}, x_1, x_{2q})
  ~~~(\xv \in {\cal T}).
 \label{Zn_mirror}
 \end{eqnarray}
It is convenient to start the proof
from the following expression of the edge Boltzmann factors
generalizing equation (\ref{generalxn}),
\begin{eqnarray}
  x_0& =& \sum_l p_l\, e^{2 \, q \,K_l}
    \nonumber\\
  x_1&=&\sum_l p_l\, e^{2 \, (q-1)K_l}  \nonumber\\
   x_2&=&\sum_l p_l\, e^{2\, (q-2)K_l}   \nonumber\\
    & &\vdots  \label{general_edgeBF} \\
   x_{2\, q-2}&=&\sum_l p_l\, e^{-2 \, (q-2)\, K_l} \nonumber\\
   x_{2\, q-1}&=&\sum_l p_l\, e^{-2 \, (q-1) \, K_l}\nonumber\\
  x_{2\, q}&=&\sum_l p_l\, e^{-2 \, q \, K_l} , \nonumber
 \end{eqnarray}
where the sum runs over $l=1, 2$ with $p_1=p, p_2=1-p$ and $K_1=K, K_2=-K$
for the $\pm J$ Ising model.
We would like the $x_k$ with $\,k$ even to remain invariant and 
the  $x_k$ with $\, k$  odd to be changed into $\,x_{2\, q \, -k}$.
As one can see from equation (\ref{general_edgeBF}),
this may be considered to be a transformation $\, K \rightarrow -K$ 
for odd $\, k$ only:
\begin{equation}
  x_k(K)\,\, \rightarrow  \,\,\, x_k((-1)^k \cdot K).
   \label{xm_sign_change}
\end{equation}
The sign $\, (-1)^k\, $ actually corresponds to the sign 
of $\, \Sigma_i \cdot \Sigma_j\, $ for neighbouring sites $i$ and $j$, where
$\, \Sigma_i =\sigma_i^{(1)}\,\sigma_i^{(2)} \, \cdots \sigma_i^{(2q)}$
and  $\, \Sigma_j  = \sigma_j^{(1)}\,\sigma_j^{(2)} \, \cdots \sigma_j^{(2q)}$,
because $\, \Sigma_i \cdot \Sigma_j\, $ is $1$ for even number of antiparallel pairs
and is $-1$ otherwise.
The formula (\ref{general_edgeBF}) can be written as
\begin{equation} 
 x_k\, = \, \, \sum_{l} p_{l} \,  
 \exp\left( K_{l} \sum_{\alpha =1}^{n} \sigma_i^{(\alpha)} \, \sigma_j^{(\alpha)}\right),
\end{equation}
where $k$ pairs among $2q$ pairs of neighbouring spins are antiparallel
$\sum_{\alpha} \sigma_i^{(\alpha)}\sigma_j^{(\alpha)}=2q-2k$.
Changing $\, K_{l}\, $ as
\begin{equation}
  K_{l}\, \, \rightarrow  \,\,\,  \Sigma_i \cdot \Sigma_j 
  \cdot  K_{l}\,
 \end{equation}
amounts to changing the edge Boltzmann factor as follows :
\begin{eqnarray}
 x_k \rightarrow  \,\,\, \sum_{l} p_{l} \,
 \exp\left(  K_{l} \cdot \Sigma_i \cdot \Sigma_j \sum_{\alpha =1}^{n}
 \sigma_i^{(\alpha)} \, \sigma_j^{(\alpha)}\right) .
\end{eqnarray}
In other words this is equivalent to performing for each replica
a non-trivial Mattis transformation depending on all the other replicas:
\begin{eqnarray}
 \sigma_i^{(\alpha)} \, \, \rightarrow  \,\,\, \tau_i^{(\alpha)} \, 
 = \, \, \Sigma_i \cdot
 \sigma_i^{(\alpha)}\, = \, \, 
 \prod_{\beta \ne \alpha} \sigma_i^{(\beta)}\, .
\end{eqnarray}
Note that, as far as dummy variables to be summed 
on are concerned, the $\, \tau_i^{(\alpha)}$'s are as good as
the $\,  \sigma_i^{(\alpha)}$'s : This is a one-to-one change of variables.
Actually one goes back from the $\, \tau_i^{(\alpha)}$'s to the  $\,  \sigma_i^{(\alpha)}$'s
by performing the same transformation as the one defining 
the $\, \tau_i^{(\alpha)}$'s :
\begin{eqnarray}
\tau_i^{(\alpha)} \, \, \rightarrow  \,\,\, \sigma_i^{(\alpha)} \, 
 = \, \, 
\prod_{\beta \ne \alpha} \tau_i^{(\beta)}\, \,\,
= \, \, \, \prod_{\beta \ne \alpha} (\Sigma_i \cdot
\sigma_i^{(\beta)}) \, \,\,
= \, \, \,  \Sigma_i^{2q-1} \cdot \Sigma_i \cdot 
\sigma_i^{(\alpha)}.
\end{eqnarray}
We have used the identity $ \, \Sigma_i^{2q}\, = \, +1$.
Therefore this Mattis transformation is an involution, and we have proved equation
(\ref{xm_sign_change}) which is equivalent to the symmetry (\ref{Zn_mirror}).

Let us point out another interesting symmetry of the partition function
which can be derived from a combination of symmetries discussed so far.
If we denote the duality transformation (\ref{x_star}) symbolically as $D$
and the exchange of $x_k$ and $x_{2q-k}$ for even $n$ and odd $k$ in equation (\ref{Zn_mirror})
as $M$, the combination
 \begin{equation}
   D_2=D \cdot M
   \label{D2}
 \end{equation}
also leaves the partition function invariant as long as the system satisfies
the conditions for $D$ and $M$ to be the true symmetry such as the
lattice structure (square lattice) and even $n$.
This duality-like symmetry under the transformation $D_2$ will be used in the next subsection.
It is worth noticing here that $D$ and $M$ commute for $n$ even ($D\cdot M=M\cdot D$), and
$D$, $M$ and $D_2$ are all involutions, $D^2 =M^2 =D_2^2={\bf 1}$.
\subsection{Subvariety and duality}\label{sec:subvariety}
The $\pm J$ Ising model with quenched randomness has remarkable properties
along a line (curve) in the phase diagram defined by the relation
 \begin{equation}
   \exp (-2K)=\frac{1-p}{p}
   \label{N_cond}
 \end{equation}
known as the Nishimori line (NL) \cite{Nishimori81,Nishimori01}.
Similar interesting behaviour is observed on the same line also
in the replicated system with $n=2$ \cite{Georges}.
In this subsection we analyze symmetries of the replicated system
with general integer $n$ under the condition (\ref{N_cond})
using the results of previous subsections.

Let us consider the $\pm J$ Ising model on the square lattice whose
parameters satisfy equation (\ref{N_cond}).
From equations (\ref{generalxn}) and (\ref{N_cond}),
the latter being equivalent to $p=w_0/(w_0+w_1)$ and $1-p=w_1/(w_0+w_1)$,
it is straightforward
to see that the following relations hold:
 \begin{equation}
  x_1=x_n,~x_2=x_{n-1},~x_3=x_{n-2},\cdots , x_k=x_{n-k+1},\cdots .
   \label{N_subvariety}
 \end{equation}
The condition (\ref{N_cond}) reduces the degree of freedom from two ($p$ and $K$)
to one.  This means that we restrict ourselves to a one-dimensional curve
in the $(n+1)$-dimensional space ${\cal S}$.
It will be useful to relax this constraint and consider the $([(n+1)/2]+1)$-dimensional
submanifold of ${\cal S}$ specified only by equation (\ref{N_subvariety}),
where $[x]$ stands for the largest integer not exceeding $x$.
We shall call this submanifold the subvariety $\cal N$:
 \begin{equation}
   {\cal N}=\{ \xv \in {\cal S}|x_k=x_{n-k+1}~(k=1,2,\cdots ,n)\}.
 \end{equation}
The subvariety $\cal N$ of course
includes the one-dimensional curve NL specified by equation (\ref{N_cond}): NL $\subset {\cal N}$.

By the duality (\ref{x_star}), $\cal N$ is transformed
into the dual ${\cal N}^*$ satisfying
 \begin{equation}
   x_1 =x_2,~x_3=x_4,~x_5=x_6,\cdots  ,x_{2m-1}=x_{2m},\cdots .
    \label{N_dual}
 \end{equation}
To prove this fact, we first note that the condition (\ref{N_subvariety})
implies that the coefficient of $x_k (=x_{n-k+1})$ on the right-hand side of
equation (\ref{x_star}) is
 \begin{equation}
   D_m^k (n)+D_m^{n-k+1}(n),
 \end{equation}
where we have written the $n$-dependence of $D_m^k$ explicitly.
Then the relation (\ref{N_dual}) follows if we can show
 \begin{equation}
   D_{2m-1}^k(n)+D_{2m-1}^{n-k+1}(n) =D_{2m}^k(n)+D_{2m}^{n-k+1}(n),
   \label{D_rel1}
 \end{equation}
the left-hand side of which is the coefficient of $x_k(=x_{n-k+1})$ of $2^{n/2}x_{2m-1}^*$
in equation (\ref{x_star}) and the right-hand side is for
$x_k(=x_{n-k+1})$ of $2^{n/2}x_{2m}^*$.
Equation (\ref{D_rel1}) can be proved by induction with respect to $n$:
The validity for small $n(=1,2,3)$ is checked trivially.
Let us assume that that equation (\ref{D_rel1}) is valid for $n$.
Then it is not difficult to show that
the same equation holds for $n+1$ using the following recursion relation
 \begin{equation}
  D_m^k(n+1)=D_m^k(n)+D_m^{k-1}(n),
  \label{D-recursion1}
 \end{equation}
which is derived by multiplying both sides of equation (\ref{t-expansion}) by $1+t$
(which amounts to the change $n\to n+1$). This ends the proof.

Comparison of equations (\ref{N_subvariety}) and (\ref{N_dual}) suggests that
the subvariety $\cal N$ is in general not self-dual, ${\cal N}\ne {\cal N}^*$.
When $n$ is even $n=2q$, the partition function has an additional symmetry
(\ref{Zn_mirror}), $M$ in the notation of equation (\ref{D2}).
Then the combined transformation $D_2=D\cdot M$ keeps the subvariety
invariant, ${\cal N}={\cal N}^*$:
The application of $D_2$ to $\cal N$ is shown to yield the same
relation as equation (\ref{N_subvariety})
 \begin{equation}
   x_1^*=x_{2q}^*,~x_2^*=x_{2q-1}^*,~x_3^*=x_{2q-2}^*,\cdots ,x_m^*=x_{2q-m+1}^*,\cdots  .
    \label{N_dual2}
 \end{equation}
The proof is outlined in \ref{sec:AppendixA}.
We therefore conclude that the subvariety $\cal N$ is {\em globally self-dual} when $n$
is even, that is, any point in the submanifold defined by equation (\ref{N_subvariety})
is mapped by $D_2$ to another point in the same submanifold
(but is not fixed point-by-point in general).
\subsection{Inversions}
The edge Boltzmann matrix $A_n$ with generic values of the elements
$(x_0, x_1, x_2,\cdots ,x_n)$ has a remarkable property that its inverse matrix
has the same structure;
the inverse  matrix has the same arrangement of elements as the original matrix.
To show this we first note that
the entries of the Boltzmann matrix 
are defined up to a common multiplicative factor, and therefore
we can also discuss, instead of the matrix inversion  
$A_n \rightarrow \,A_n^{-1}$,  a homogeneous
matrix inversion transformation $\, I$ (actually a homogeneous
polynomial transformation):
\begin{eqnarray}
 A_n\, \rightarrow \,
 A_n^{-1} \cdot \det(A_n) .
 \label{homoI}
\end{eqnarray}
Let us also introduce the homogeneous transformation
$\, J$ corresponding 
to the inversion of elements of the dual matrix (Hadamard inverse),
$\, y_k \, \rightarrow \, 1/y_k\,  (k \, = \, 0, \, 1, \, \cdots , n)$
\footnote{
Transformation
$\, J$ is analogous to negating various coupling constants.
},
\begin{eqnarray}
 J\, : \qquad y_k \, \quad \longrightarrow \quad 
 \prod^{m=n}_{m=0, m \ne k} y_m .
\end{eqnarray}

At first sight, performing the matrix inversion of
the  $\, 2^n \times 2^n \, $  Boltzmann  matrix
$\, A_n\, $ may seem to yield quite large calculations. 
However, since the duality transformation 
actually diagonalizes matrix $\, A_n$ as was mentioned in relation to
equation (\ref{general_diagonal}),
it is straightforward to see that the matrix inversion 
{\em just amounts to changing the dual variables $\, x_m^{*}$ into 
their simple inverse} : $\, x_m^{*} \,\rightarrow \, 1/x_m^{*}$.
From this remark, it is straightforward to see
that the matrix inverse of $\, A_n\, $
is a $\, 2^n \times 2^n \, $ matrix of the same form as the original
$\, A_n$ (but of course with 
different elements : $\, x_k \,\rightarrow \, x'_k$). 
With obvious notations, and since there is no possible confusion, 
we will also denote by $\, I$ 
this transformation on the $\, x_k$'s :
\begin{eqnarray}
I \, : \qquad x_k \, \quad \longrightarrow \quad \, x'_k
\end{eqnarray}
Up to a common multiplicative factor one thus has (since $\, D$ is an
involution, that is, $\, D^2 \, = \,{\bf 1}$) :
\begin{eqnarray}
\label{DID}
 D \cdot I \cdot D \, \propto \, \,  J, \qquad \hbox{or} 
\qquad
I \propto \, \, D \cdot J \cdot D .
\end{eqnarray}

{\bf Remark :} One should recall that the two involutions $\, I\,$ and $\, J\,$ 
are {\em actually (non-linear) symmetries} of the phase diagrams of (anisotropic) 
spin edge lattice 
models~\cite{dAuriac,MeAnMaRo94,baxt}. From equation (\ref{DID}) one sees that
these ``non-linear'' symmetries of the phase diagrams are closely
related to the (linear) duality symmetry. A duality symmetry exists
when the edge Boltzmann matrix corresponds to cyclic matrices
or semi-direct product of cyclic matrices~\cite{Wu-Wang}. However,
when a {\em duality symmetry does not exist}, like for instance the Ising
model in a magnetic field, the two ``non-linear'' symmetries
$\, I\,$ and $\, J\,$ {\em still exist and can still be used to
analyse the phase diagram}~\cite{MeAnMaRo94}.

%
\subsection{Spin representation of the dual Boltzmann factor}\label{sec:DualBF}

The elements of the dual edge Boltzmann factors of the replicated $\pm J$
Ising model (\ref{xmstar_pmJ}) have an
interesting symmetry.  It is instructive to take the ratios of $x_1^{*}, x_2^{*},\cdots $ to
$x_0^{*}$ (which is equivalent to setting the energy level of all-parallel state to 0):
\begin{equation}
\begin{array}{rcl}
 \displaystyle
   x_{2m-1}^*/x_0^* &=&\displaystyle(2p-1) \left(\frac{w_0-w_1}{w_0+w_1}\right)^{2m-1}
     =(2p-1)\tanh^{2m-1} K \\
   \displaystyle x_{2m}^*/x_0^*  &=&\tanh^{2m} K .
  \end{array}
  \label{dual_BF5}
\end{equation}
It is observed that the right-hand side is multiplied by $\tanh K$
each time the number of antiparallel spin pairs $m$ increases.
Another factor $2p-1$ appears alternately.
Therefore the right-hand side of equation (\ref{dual_BF5}) can be expressed
as a simple Boltzmann factor of the dual system,
 \begin{equation}
   A \exp \left\{ K^{*} (S^{(1)}+S^{(2)}+\cdots +S^{(n)})+K_p^{*}
   S^{(1)}S^{(2)}\cdots S^{(n)}\right\},
   \label{dualBF0}
 \end{equation}
where $S^{(\alpha )}$ is the product of neighbouring dual spins in the $\alpha$th replica
$(S^{(\alpha)}=S_i^{(\alpha)}S_j^{(\alpha)})$, and
$K^{*}$ and $K_p^{*}$ are the dual couplings corresponding to thermal and randomness
parameters:
 \begin{equation}
   \tanh K=e^{-2K{*}},~~2p-1 =e^{-2K_p^{*}}.
 \end{equation}
This expression (\ref{dualBF0}) has a very interesting interpretation.
The first part can be interpreted as being driven by thermal fluctuations
since it has only the (dual) thermal coupling $K^*$ in front of the spin variables.
This first term is decoupled explicitly from
the second part which is understood to be driven by quenched randomness as it
is controlled by the (dual) coupling $K_p^*$ determined only by $p$.
The first term causes ferromagnetic ordering of dual spin variables
whereas the second term enhances spin-glass-like multi-replica ordering.

The condition (\ref{N_cond}) is written in terms of the dual couplings as $K^*=K_p^*$.
Therefore, on the NL, the above-mentioned
two types of ordering tendency exactly balance.
This reminds us of the result that the ferromagnetic ordering dominates above the
NL whereas spin-glass order is larger below \cite{Nishimori02}.
The balance $K^{*}=K_p^{*}$ may also be called enhanced symmetry because the single-replica
and the multi-replica terms have exactly the same coupling.

It should be pointed out that a larger type of enhanced symmetry is realized by the
$2^n$-state standard scalar Potts model which has the edge Boltzmann factor
\begin{eqnarray}
 & & A\exp \left\{  K_1(S^{(1)}+S^{(2)}+\cdots +S^{(n)})\right.  \nonumber\\
 && +K_2 (S^{(1)}S^{(2)}+S^{(1)}S^{(3)}
 +\cdots  +S^{(n-1)}S^{(n)})
   \nonumber\\
 &&\left. +\cdots +K_n S^{(1)}S^{(2)}\cdots S^{(n)}\right\}
 \label{dual_BF2}
\end{eqnarray}
with all couplings equal $K_1=K_2=K_3=\cdots =K_n$:
It is straightforward to confirm that this Boltzmann factor has just two values,
one for $S^{(1)}=S^{(2)}=\cdots =S^{(n)}=1$ and the other for all other
configurations, thus representing the standard scalar Potts model.

These two models, the $\pm J$ Ising model (\ref{dualBF0}) on the NL (with $K^*=K_p^*$) and 
the $2^n$-state standard scalar Potts model (\ref{dual_BF2})
(with $K_1=K_2=K_3=\cdots =K_n$), coincide when $n=2$ but not in general.
We may therefore regard the $n$-replicated $\pm J$ Ising model on the NL
as a system with similar but a little weaker symmetry than the standard scalar Potts model
with $2^n$ states.
Another point to notice is that the original Boltzmann factors
(\ref{generalxn}) of the $\pm J$ Ising model
are not expressible in a simple form like equation (\ref{dualBF0}):
If we try to do so by adjusting the coupling constants in equation (\ref{dual_BF2}),
each $K_i$ will appear with a different value from the other coefficients.
Only the dual system has the simple form of the edge Boltzmann factor
($K_2=K_3=\cdots =K_{n-1}=0$).

\subsection{Results derived from the dual Boltzmann factor}\label{sec:IneqBF}

\begin{figure}[hbt]
  \begin{center}
  \includegraphics[width=0.58\linewidth]{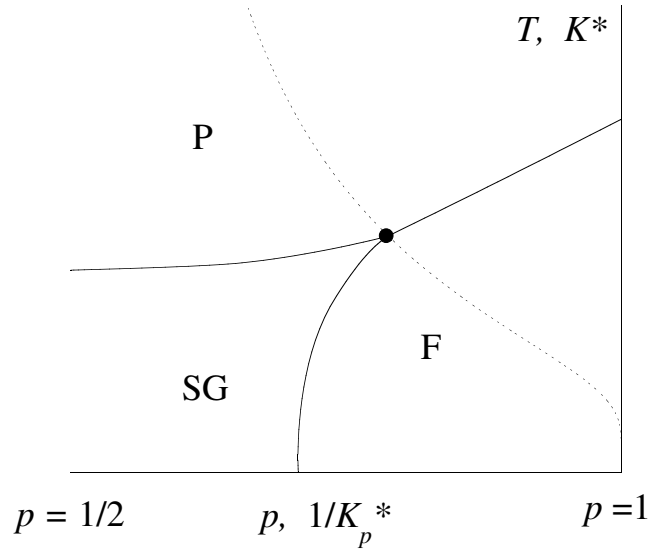}
  \end{center}
 \caption{Generic phase diagram of the $n$-replicated Ising model.
 The multicritical point (black dot) lies on the NL (shown in a thin dotted curve).
 }
 \label{fig:generic-phase-diagram}
\end{figure}

The effective edge Boltzmann factor (\ref{dualBF0}) shows that
the replicated $\pm J$ Ising model in the dual representation has a relatively simple
Hamiltonian with positive ferromagnetic couplings.
This fact enables us to analyze the model using known results on
ferromagnetic spin systems.
One of our interests to be discussed later will be the structure of the phase diagram
of the $n$-replicated system.
The phase diagram drawn in terms of the original parameters $p$ and $T$
should look topologically the same as the one with axes $1/K_p^{*}$ and $K^{*}$
because $1/K_p^{*}$ is a monotone increasing function of $p$
and $K^{*}$ is also monotonic in $T$ (see figure \ref{fig:generic-phase-diagram}).
Note that in the present paper we call the point marked in black dot in figure
\ref{fig:generic-phase-diagram}
the multicritical point irrespective of the order of transition.
Of course one should take into account that ordered and disordered
phases are exchanged between the original and dual systems.

We first check a few limiting cases.  According to
the dual Boltzmann factor (\ref{dualBF0}),
when $K_p^{*} \to 0$ ($p\to 1$), the system decouples into
$n$ independent ferromagnetic Ising models as expected.
If, on the other hand, $K^{*}\to 0$ (or the low-temperature limit $T\to 0$
in the original variable), only the multi-replica
coupling $K_p^{*}$ survives in equation (\ref{dualBF0}).
By redefining spins as $\sigma_i =S_i^{(1)}S_i^{(2)}\cdots S_i^{(n)}$,
the system reduces to a simple ferromagnetic Ising model.
The critical point exists at $K_p^{*}=K_{\rm c}^{\rm F}$, where $K_{\rm c}^{\rm F}$ is the
critical point of the ferromagnetic Ising model satisfying $e^{-2K_{\rm c}^{\rm F}}=\sqrt{2}-1$.
Therefore the original system in the ground state has a critical point
at $p_{\rm c}=\frac{\sqrt{2}}{2}=0.7071$ (derived from $e^{-2K_p^*}=\sqrt{2}-1$)
irrespective of $n$ as long as $n$ is a positive integer.
Note that the quenched limit $n\to 0$ is
expected to have a different critical probability $p_{\rm c}$ near 0.89 (see the next section).
This observation means that the two limits $n\to 0$ and $T\to 0$ do not commute.

The other limit $K_p^{*}\to \infty$
corresponds to $p=\frac{1}{2}$ in the original variable.
According to equation (\ref{dualBF0}), the dual spins are then constrained
as $S^{(1)}S^{(2)}\cdots S^{(n)}=1$.
It follows that the $n$th spin variable can be expressed as
$S^{(n)}= S^{(1)}S^{(2)}\cdots S^{(n-1)}$.
Then the dual Boltzmann factor becomes
 \begin{equation}
A \exp \left\{ K^{*} (S^{(1)}+S^{(2)}+\cdots +S^{(n-1)}+S^{(1)}S^{(2)}\cdots S^{(n-1)})\right\}.
   \label{dualBF5}
 \end{equation}
This is exactly the Boltzmann factor of the $(n-1)$-replicated system
with $K_p^{*}=K^{*}$.
We have therefore established the following relation for the partition functions
of the $n$- and $(n-1)$-replicated $\pm J$ Ising models:
 \begin{equation}
   Z_n(K,K_p=0) \propto Z_{n-1}(K,K_p=K),
    \label{ZnZn-1}
 \end{equation}
where the partition function is expressed by using the original couplings
($K_p$ defined by $e^{-2K_p^{*}}=\tanh K_p$ is the dual of $K_p^{*}$).
The trivial constants in front of the partition function are omitted.
The identity (\ref{ZnZn-1}) proves that the $n$-replicated system with
$p=\frac{1}{2}~(K_p=0)$ is equivalent to the $(n-1)$-replicated system on the NL, $K=K_p$
(equivalent to $K^*=K_p^*$).

The effective edge Boltzmann factor (\ref{dualBF0}) represents a system
with ferromagnetic couplings only.
Thus the Griffiths inequalities hold \cite{Griffiths}.
In particular, first derivatives of an arbitrary correlation function
are positive semi-definite:
 \begin{equation}
 \frac{\partial}{\partial K^{*}}\langle S_i^{(\alpha)}S_j^{(\beta)}S_k^{(\gamma)}\cdots \rangle
   \ge 0,~~~
  \frac{\partial}{\partial K_p^{*}}\langle S_i^{(\alpha)}S_j^{(\beta)}S_k^{(\gamma)}\cdots \rangle
   \ge 0.
    \label{Griffiths}
 \end{equation}
These inequalities imply that the boundaries between ferromagnetic
and non-ferromagnetic phases are monotonic in $K^{*}$ and $K_p^{*}$
(and in the original variables $T$ and $p$)
as drawn schematically in figure \ref{fig:generic-phase-diagram}.
The reason is that, otherwise, the system will show reentrant behaviour
and consequently the order parameters will be non-monotonic, violating (\ref{Griffiths}).

An interesting consequence of monotonic behaviour of the phase boundaries
is that the transition temperature at $p=\frac{1}{2}$ between the paramagnetic
and spin glass phases (to be denoted as
$T_{\rm c}^{(n)}(\frac{1}{2})$) is smaller than or equal to the temperature of the
multicritical point $T_{\rm c}^{(n)}({\rm MCP})$ which lies generically on the NL.
Since we have already established in equation (\ref{ZnZn-1}) that the $n$-replicated system on the line $p=\frac{1}{2}$ is
equivalent to the $(n-1)$-replicated system on the NL,
it is concluded that 
 \begin{equation}
   T_{\rm c}^{(n-1)}({\rm MCP}) = T_{\rm c}^{(n)} \left(\frac{1}{2}\right) \le T_{\rm c}^{(n)}({\rm MCP}).
   \label{Tcineq}
 \end{equation}
In terms of $p$, this inequality reads, using $K=K_p$ ($\tanh K=2p-1$)
satisfied by the multicritical point to rewrite $T$ by $p$,
 \begin{equation}
  p_{\rm c}^{(n-1)}({\rm MCP}) \ge p_{\rm c}^{(n)}({\rm MCP}).
   \label{pc-bound}
 \end{equation}
The value of $p$ at the multicritical point is a monotone decreasing function of $n$.
We can then derive a lower bound for $p_{\rm c}^{(0)}({\rm MCP})$
of the quenched system $n\to 0$ if we remember that the $n=1$ annealed
system is easily solved by reducing it to the regular ferromagnetic Ising model.
The result is 
 \begin{equation}
   p_{\rm c}^{(0)}({\rm MCP}) \ge \frac{1+\sqrt{\sqrt{2}-1}}{2}=0.821797....
   \label{pc-bound2}
 \end{equation}
Although this bound is not very tight since the expected value of
the left-hand side is approximately 0.89 as discussed in the next section,
it is nevertheless useful to have a mathematically rigorous bound
(if we accept that the $n\to 0$ limit causes no problems).

We can also derive an inequality
 \begin{equation}
   T_{\rm c}^{(n)} \left(\frac{1}{2}\right) \le T_{\rm c}^{(n+1)} \left(\frac{1}{2}\right)
 \end{equation}
from equation (\ref{Tcineq}).

Another non-trivial fact is the absence of reentrant behaviour as
mentioned already.
It is well established for $n=1$ and $n=2$ that
the boundary between the ferromagnetic and non-ferromagnetic phases
below the multicritical point is never reentrant as depicted in
figure \ref{fig:generic-phase-diagram}
in accordance with the present rigorous result \cite{Georges}.
It is a subtle matter whether or not this monotonicity still holds
in the quenched limit $n\to 0$:
If this is the case, the phase boundary below the multicritical point
should be a vertical straight line in accordance with the argument in \cite{Nishimori86}
because the absence of ferromagnetic
phase to the left of the multicritical point at any temperature is rigorously established
in the quenched model \cite{Nishimori81}.
However we avoid to make a definite statement here
since it is not obvious that monotonicity derived from
inequalities proved for positive integer $n$ remains valid in the limit $n\to 0$.

\subsection{Relations between $n$ and $n+1$}\label{sec:gauge}

It has been established in equation (\ref{ZnZn-1}) that the $n$-replicated system with $p=\frac{1}{2}$
is equivalent to $(n-1)$-replicated system on the NL on the
square lattice.
This result has already been pointed out by Georges \etal \cite{Georges}.
We here derive several further relations on physical quantities
of $n$- and $(n+1)$-replicated systems.
The arguments in the present subsection apply to the replicated $\pm J$ Ising
model on an arbitrary lattice in an arbitrary dimension.
No duality will be used.

It is instructive to rederive the relation between the partition functions
$Z_n$ and $Z_{n+1}$ without recourse to duality.
The randomness average of the $n$-replicated partition function
of the $\pm J$ Ising model is written as \cite{Nishimori81,Nishimori01}
 \begin{equation}
   Z_n(K, K_p)=\frac{1}{(2\cosh K_p)^{N_B}}\sum_{\tau }
     e^{K_p \sum \tau_{ij}} Z(K)^n,
 \end{equation}
where $N_B$ is the number of bonds of the lattice, $\tau_{ij}=\pm 1$ is
the sign of $J_{ij} (=\pm J)$, and $Z(K)$ is the partition function
with fixed randomness.
After gauge transformation and summation over gauge variables,
the exponential in the above expression turns to a partition function
with coupling $K_p$ \cite{Nishimori81,Nishimori01}:
 \begin{equation}
   Z_n(K, K_p)=\frac{1}{2^N (2\cosh K_p)^{N_B}}\sum_{\tau }
     Z(K_p) Z(K)^n,
 \end{equation}
where $N$ is the total number of spins.
It readily follows from this equation that $Z_n(K, K)$ and $Z_{n+1}(K, 0)$ are
essentially equal to each other:
 \begin{equation}
    2^N (2\cosh K)^{N_B} Z_n(K, K)= 2^{N+N_B} Z_{n+1}(K, 0).
    \label{ZnZn1}
 \end{equation}
This is the identity (\ref{ZnZn-1}) we already derived using duality for the square lattice.
The present result is more general as we did not use the properties
of a specific lattice.
%

Similar identities hold for order parameters. The magnetization is defined by
 \begin{equation}
  m_n(K, K_p)=\frac{\sum_\tau e^{K_p\sum \tau_{ij}}
  \left( \sum_S S_i e^{-\beta H(S)} \right)  Z(K)^{n-1}}
   {\sum_\tau e^{K_p\sum \tau_{ij}} Z(K)^{n}},
  \end{equation}
where $\beta =1/k_B T$.
Spins on boundaries are fixed to up states to avoid trivial vanishing of
the single-spin expectation value due to global inversion symmetry.
Gauge transformation brings this equation into
 \begin{equation}
  m_n(K, K_p)=\frac{\sum_\tau \left( \sum_\sigma \sigma_i 
  e^{K_p\sum \tau_{ij}\sigma_i \sigma_j} \right)
  \left(  \sum_S S_i e^{-\beta H(S)}\right) Z(K)^{n-1}}
   {\sum_\tau Z(K_p) Z(K)^{n}}.
   \label{mn}
  \end{equation}
This expression of magnetization is to be compared with the spin glass order parameter,
which can be written using gauge transformation as
\begin{eqnarray}
  q_n(K, K_p)&=&\frac{\sum_\tau e^{K_p\sum \tau_{ij}}
      \left( \sum_S S_i e^{-\beta H(S)}\right)^2 Z(K)^{n-2}}
       {\sum_\tau e^{K_p\sum \tau_{ij}} Z(K)^{n}}  \nonumber\\
   &=& \frac{\sum_\tau 
    \left( \sum_S S_i e^{-\beta H(S)}\right)^2 Z(K_p) Z(K)^{n-2}}
   {\sum_\tau Z(K_p) Z(K)^{n}}.
   \label{qn}
\end{eqnarray}
Comparison of equations (\ref{mn}) and (\ref{qn}) immediately leads to
\begin{equation}
 m_n(K, K)=q_n(K, K)=q_{n+1}(K, 0).
  \label{mnqn}
\end{equation}
The first equality shows that there is no spin glass phase on the NL
if we define the spin glass phase by $m_n=0$ and $q_n>0$.
The second equality of (\ref{mnqn}) indicates that the spin glass order parameter of the
$n$-replicated system on the NL is exactly equal to that
of the $(n+1)$-replicated system with $p=\frac{1}{2}$.

The above identity (\ref{mnqn}) on the order parameters can be generalized to a relation
between the distribution functions of order parameters \cite{Nishimori01}.
We define and gauge-transform the distribution function of magnetization as
 \begin{eqnarray}
  & &P_m^{(n)}(x; K, K_p) \nonumber \\
  &&=\frac{\sum_\tau e^{K_p\sum \tau_{ij}}\left( 
    \sum_S  \delta ( x-\frac{1}{N}\sum_i S_i )  e^{-\beta H(S)} \right)  Z(K)^{n-1}}
    {\sum_\tau e^{K_p\sum \tau_{ij}} Z(K)^{n}}
      \nonumber \\
   &&=  \frac{\sum_\tau \left( \sum_S \sum_\sigma \delta ( x-\frac{1}{N}\sum_i S_i \sigma_i)
  e^{-\beta H(S)}e^{-\beta_p H(\sigma )} \right)  Z(K)^{n-1}}
   {\sum_\tau Z(K_p) Z(K)^{n}},
   \label{Pm}
  \end{eqnarray}
where $\beta_p H(\sigma)$ is the Hamiltonian with coupling $K_p$ and
spin variables $\sigma$.
The distribution of spin glass order parameter, measuring the overlap
of two replicas, is treated similarly.
Using gauge transformation it is written as
 \begin{eqnarray}
 & & P_q^{(n)}(x; K, K_p) \nonumber \\
 &&=\frac{\sum_\tau e^{K_p\sum \tau_{ij}}
    \left( \sum_S  \sum_\sigma \delta ( x-\frac{1}{N}\sum_i S_i \sigma_i)  
      e^{-\beta H(S)} e^{-\beta H(\sigma )} \right)  Z(K)^{n-2}}
    {\sum_\tau e^{K_p\sum \tau_{ij}} Z(K)^{n}}
      \nonumber \\
   &&=  \frac{\sum_\tau \left( \sum_S \sum_\sigma \delta ( x-\frac{1}{N}\sum_i S_i \sigma_i)
  e^{-\beta H(S)}e^{-\beta H(\sigma )} \right)  Z(K_p) Z(K)^{n-2}}
   {\sum_\tau Z(K_p) Z(K)^{n}}.
   \label{Pq}
  \end{eqnarray}
It is easy to see from equations (\ref{Pm}) and (\ref{Pq})
that the following relation holds
 \begin{equation}
   P_m^{(n)}(x; K, K)=P_q^{(n)}(x; K, K)=P_q^{(n+1)}(x; K, 0).
    \label{PmPq}
 \end{equation}
Since the distribution function of a single-replica variable
like magnetization $P_m(x)$ has only a trivial structure (with at most
two delta peaks), it follows from the first equality of (\ref{PmPq})
that the distribution of the spin glass order parameter $P_q(x)$ is also
trivial on the NL for any $n$:
There exists no complex structure such as replica-symmetry breaking.
The second equality of (\ref{PmPq}) then proves that the same trivial
structure holds for the spin-glass distribution function of
$(n+1)$-replicated system with $p=\frac{1}{2}$.
Thus the spin glass phase of the $(n+1)$-replicated system with $p=\frac{1}{2}$
has only a trivial structure.
We should be careful when we consider the possibility to apply this result to the
distribution function $P_q^{(n+1)}$ of the quenched system which,
in the present context, corresponds to $n+1\to 0$ or $n\to -1$.
The arguments in the present subsection are mathematically rigorous
only for positive integer $n$.

\subsection{Generalization}

Many of the arguments so far have been on the replicated $\pm J$ Ising model although some
of the symmetry relations are valid in more general systems with generic edge Boltzmann 
matrix with hierarchical structure.
In the present subsection we apply the duality relations to the random Ising model
with general coupling values as well as to the random $Z_q$ model
which includes the random chiral Potts model.

Let us first treat the $n$-replicated Ising model with a set of couplings
$K_1, K_2, K_3, \cdots$ with respective probabilities $p_1, p_2, p_3, \cdots$.
The sign of $K_j$ is arbitrary in this subsection:  All the couplings can be
ferromagnetic, for example.
The hierarchical structure of the edge Boltzmann matrix is the same as the
one already discussed for the $\pm J$ Ising model.

The edge Boltzmann factors are
 \begin{eqnarray}
   x_0&=&\sum_j p_j e^{nK_j} \nonumber\\
   x_1&=&\sum_j p_j e^{(n-2)K_j} \nonumber\\
   x_2&=&\sum_j p_j e^{(n-4)K_j}  \\
   & & \vdots \nonumber\\
   x_n&=&\sum_j p_j e^{-nK_j} .
 \end{eqnarray}
The dual are then
 \begin{equation}
  \begin{array}{rcl}
  \displaystyle
   2^{n/2}x_0^* &=& \sum_j p_j (e^{K_j}+e^{-K_j})^n \\
   2^{n/2}x_1^* &=& \sum_j p_j (e^{K_j}+e^{-K_j})^{n-1} (e^{K_j}-e^{-K_j}) \\
   2^{n/2}x_2^* &=& \sum_j p_j (e^{K_j}+e^{-K_j})^{n-2} (e^{K_j}-e^{-K_j})^2 \\
    & \vdots &\\
   2^{n/2}x_n^* &=& \sum_j p_j (e^{K_j}-e^{-K_j})^n,
  \end{array}
 \end{equation}
a generalization of equations (\ref{xms}) and (\ref{xmstar_pmJ}).
It is sometimes convenient to take the continuum limit of randomness distribution.
The original and dual Boltzmann factors are then
\begin{eqnarray}
 x_k &=& \int du\, P(u)\,  e^{(n-2k)\beta u} 
   \label{xk}\\
 2^{n/2} x_m^* &=& \int du\, P(u)\, (e^{\beta u}+e^{-\beta u})^{n-m} (e^{\beta u}-e^{-\beta u})^m.
   \label{xdk}
\end{eqnarray}

A generalization of equation (\ref{N_cond})
is expressed as the probability distribution of the form \cite{Nishimori81,Nishimori01}
 \begin{equation}
   P(u)=e^{\beta u}F(u^2),
   \label{generaln}
 \end{equation}
where $F(u^2)$ is an arbitrary even function of $u$ and should be chosen to
satisfy the normalization condition of $P(u)$.
The constraint (\ref{generaln}) applied to the edge Boltzmann factors (\ref{xk})
and (\ref{xdk}) gives the symmetry properties
 \begin{eqnarray}
 && x_1=x_n,~~x_2=x_{n-1},~~x_3=x_{n-2},\cdots \label{x1}
 \\
 &&x_1^{*}=x_2^{*},~~x_3^{*}=x_4^{*},~~
  x_5^{*}=x_6^{*},\cdots .\label{xs1}
 \end{eqnarray}
The first equation (\ref{x1}) can be checked directly from equations (\ref{xk})
and (\ref{generaln}).  The second (\ref{xs1}) has already been proved
generically in equation (\ref{N_dual}).
When $n$ is even, a further symmetry exists as in equation (\ref{N_dual2}),
implying that the subvariety $\cal N$ is globally self-dual, ${\cal N}={\cal N}^*$.

We next turn to the random chiral $Z_q$ model for which
again we use the Wu-Wang duality.
The arguments in the previous subsections do not apply directly because we have used
a binary (Ising) structure of basic variables.
Each spin variable is now assumed to have $q$ components to be symbolized by integers from
$0$ to $q-1$.
The interaction will be written as $V(k)$ when the difference in the
neighbouring spin variables is $k$.
The Ising model has $q=2$, and the $q$-state standard scalar
Potts model satisfies $V(1)=V(2)=\cdots =V(q-1)$.
The interaction is cyclic, $V(k+q)=V(k)$.
Chiral randomness is assumed to exist: The interaction energy for the difference
in neighbouring spin variables being $k$ is changed
from the non-random value $V(k)$ to $V(k+l)$ with probability $p_l$.

The $k$th edge Boltzmann factor of the original lattice is
 \begin{equation}
   x_k=\sum_{l=0}^{q-1} p_l\, e^{n V(l+k)}.
   \label{xkq}
 \end{equation}
The dual Boltzmann factors are obtained by Fourier transform of the edge Boltzmann
matrix, a $q$-state generalization of equation (\ref{general_diagonal}),
and we only write the result here.
See \ref{sec:AppendixB} for some details.

 \begin{eqnarray}
  q^{n/2} x_0^{*} &=& \sum_l p_l \left( \sum_{k=0}^{q-1} e^{V(k+l)} \right)^n 
   \label{xkqd}\\
  q^{n/2} x_1^{*} &=& \sum_l p_l \left( \sum_{k} e^{V(k+l)} \right)^{n-1}
        \left( \sum_k \omega^k e^{V(k+l)} \right)\\
  q^{n/2} x_2^{*} &=& \sum_l p_l \left( \sum_{k} e^{V(k+l)} \right)^{n-1}
        \left( \sum_k \omega^{2k} e^{V(k+l)} \right)\\
  &&\vdots \nonumber\\
  q^{n/2} x_{q-1}^{*} &=& \sum_l p_l \left( \sum_{k} e^{V(k+l)} \right)^{n-1}
        \left( \sum_k \omega^{(q-1)k} e^{V(k+l)} \right)\\
  q^{n/2} x_{(q-1)+1}^{*} &=& \sum_l p_l \left( \sum_{k} e^{V(k+l)} \right)^{n-2}
        \left( \sum_k \omega^k e^{V(k+l)} \right)^2 \\
  && \vdots \nonumber
\end{eqnarray}
where $\omega$ is the $q$th root of unity.
These expressions, in particular those for $x_0$ and $x_0^{*}$, will
be used in later sections.

\section{Complexity generated by inversions}\label{sec:complexity}
Since $\, I$ and $\, J$ are two (non-linear) symmetries 
of the phase diagram of the (anisotropic) 
model \cite{dAuriac,MeAnMaRo94,baxt}, it is natural to combine these two involutions considering
the {\em infinite order} (birational) transformation 
 $\, K \, = \, \, I \cdot J$. 
 Keeping in mind that 
$\, I$ is basically $\, D \cdot J \cdot D$ as in equation (\ref{DID}),
we find that the iteration of  $\, K \, = \, \, I \cdot J$
amounts to iterating $\, k \, = \, \, D \cdot J$.

The transformation $\, K$ is
 generically an {\em infinite order}
(birational) transformation which is 
a {\em canonical symmetry} of the
parameter space of the model and an infinite discrete symmetry
of the Yang-Baxter equations
(star-triangle relations for spin-edge models)
 when the model is integrable~\cite{baxt}.
This amounts to studying the iteration of transformation $\, K$
and, in particular, the {\em complexity}
 of this transformation~\cite{complex}, namely
the {\em growth of the degrees}~\cite{zeta} 
of the (numerator or denominator of the)
successive rational expressions encountered in the
iteration\footnote{Or the purely numerical growth rate of the number of
  digits of the (numerators or denominators of the) 
successive rational numbers obtained when iterating
an initial rational point~\cite{dAuriac}. Since the entries
of the matrices grow exponentially, one has to use special
representations of the integers allowing to manipulate 
large values~\cite{gnu}}.  Generically
 one gets an {\em exponential 
growth of the calculations} like $\, \lambda^L \, $ with the number $\, L$ 
of iterations. We will call from now on 
$\, \lambda$ {\em the complexity of the transformation $\, K$ 
or the complexity of the associated spin edge lattice model}.
In this ``complexity'' framework, an enhanced symmetry,
occurring for some submanifold (in fact an algebraic subvariety)
 of the parameter space of the model (here the $\, x_k$'s), 
corresponds to 
a {\em reduction of the complexity} 
$\, \lambda\, $ to a {\em smaller 
value} : the integrable subcases pop out as algebraic 
subvarieties 
associated with a (at most) {\em polynomial growth} of the
calculations~\cite{dAuriac,baxt} ($\lambda \, = \, 1$).

Let us first remark that the subvariety
 ${\cal N}$ 
is clearly  (globally) invariant 
 by $\, J$. The duality $\, D$ 
permutes  ${\cal N}$ with its dual subvariety ${\cal N}^{*}$.
The dual subvariety ${\cal N}^{*}$ (see equation (\ref{N_dual})) 
is also clearly (globally) invariant 
 by $\, J$. Keeping in mind that 
$\, I$ is basically $\, D \cdot J \cdot D$, one deduces 
immediately that  ${\cal N}$ is also  {\em (globally) invariant 
} by $\, I$, and is, thus, is {\em (globally) invariant  by}  
$\, K \, = \, \, I \cdot J$ (${\cal N}$ is (globally) invariant  by
 $\,  k^2 \,$ where
$\, k \, = \, \, D \cdot J$). 

We have calculated the ``complexities'' $\, \lambda\, $ corresponding
to the iteration of $\, k \, = \, D \cdot J$, 
for a  general $\, 2^n \times 2^n $ {\em cyclic} matrix,
then for the general $\, 2^n \times 2^n $ 
matrix (\ref{general_BM}) for various number of replicas $\, n$
and, finally, for Boltzmann factors restricted to 
the subvariety
${\cal N}$ of the $\, 2^n \times 2^n $ 
matrices (\ref{general_BM}) . The results are summarized in the
following table :

\vskip .5cm 
{\centering \textbf{\begin{tabular}{|c|c|c|c|c|}
\hline 
\( n \)&
$n$=2&
$n$=3&
$n$=4&
$n$=5\\
\hline
\hline 
\hline 
Cyclic &1
&5.8284
&13.9282
&29.9666
\\
\hline 
Matrix (\ref{general_BM}) &1
&2
&3.2143
&4.2361
\\
\hline 
${\cal N}$ &1
&1.4142
&1.6180
&2.3344
\\
\hline 
\end{tabular}\label{table1}}\par}
\vskip .7cm
These values are all related to the solutions
of polynomial equations with integer coefficients: They are
the inverse of the smallest root of various polynomials
 with integer coefficients. For instance 
the complexities of a general $\, 2^n \times 2^n $ {\em cyclic} matrix, 
displayed in the first row of this table, 
correspond to the smallest root of polynomial 
$\,\, \, 1-(2^n-2)\, t \, +t^2\,\,$
for various values of $\, n$. The complexities for generic
 $\, 2^n \times 2^n $  
matrices of the form (\ref{general_BM}), associated with $\, n$
replicas, correspond respectively to  $\, 1-\, t$, $\, 1-2\, t$,
 $\, 1-3\,t-t^2+t^3$, and $\,1-4\,t-t^2$. Finally,
the complexities for
 $\, 2^n \times 2^n $  
matrices of the form (\ref{general_BM}) {\em restricted to
the  subvariety} $\, {\cal N}$,
respectively  correspond to  $\, 1-\, t$,
$\, 1-2\,t^2$,  $\, 1-t-t^2$,
and $\,1-6\, t^2 +3\, t^4$.

We have also obtained the complexities $\lambda_K$ associated
with the iteration of $\, K \, = \, I \cdot J$, instead of $\, k\, = \, D
\cdot J$. These $\lambda_K$'s are (as it should, since 
$\, K \,$ is equivalent to $\, k^2$) the {\em  square} of the
 $\lambda$'s displayed in the previous table.
For a general $\, 2^n \times 2^n$ {\em non-cyclic matrix} 
for which {\em one does not have a duality anymore},
one, surprisingly, finds, up to the accuracy of our calculations,
the {\em same complexities} $\lambda_K$
as for a  {\em cyclic} $\, 2^n \times 2^n\, $ matrix ! The
$\, 2^n \times 2^n $ matrices of the
 form (\ref{general_BM})  are, of course,
subcases of the {\em general non-cyclic} 
 $\, 2^n \times 2^n\, $ matrices.
One thus finds {\em quite a drastic complexity reduction}
 ($(5.82842)^2\, \,
\rightarrow \, \, 2^2, \, (13.9282)^2 \, \,
\rightarrow \, \, (3.214319)^2$ , ...) when restricting 
$\, 2^n \times 2^n $ matrices to the form (\ref{general_BM}) 
corresponding to effective Boltzmann matrix for $\, n$ replicas.
The replica analysis thus defines naturally a class of highly remarkable 
(non-random) lattice models, namely (\ref{general_BM}),
 which are very interesting {\em  per se}.
We also see very clearly, that condition $\, {\cal N}$
defines a  highly singled-out subvariety of the previous 
remarkable models for which further {\em drastic
 reductions of complexity occur (enhanced symmetry)}.

It would be tempting to restrict models (\ref{general_BM})  to various 
(global or point-by-point) self-dual
conditions (like $x_0^{*} \, = \, x_0$, $x_1^{*} \, = \, x_1$, ...) 
in order to see similar 
 drastic reductions of complexities, and possibly 
a polynomial growth of complexity on some integrable subvariety,
or  on some critical submanifold\footnote{
including the cases of first order transitions
}
 (possibly given by intersections
of the subvariety $\, {\cal N}$ and various self-dual
conditions) of the $\, x_0, \, \cdots x_n$ parameter space $\cal S$.
Such an analysis, in fact, requires, on a square lattice
for instance, to generalize our isotropic 
effective model (\ref{general_BM}) to an {\em anisotropic} effective model 
with {\em two different sets of horizontal and vertical Boltzmann matrices}
both of the form (\ref{general_BM}). The calculations become much more
subtle and involved and, thus, will be detailed elsewhere. 
Let us briefly say that our preliminary ``complexity analysis'' 
results are not 
in disagreement with the numerical results displayed below.

\section{Conjecture on the location of the multicritical point}\label{sec:MCP}
We have discussed symmetry and complexity of replicated random
spin systems and their generalizations.  We now apply the results 
to the presentation of a conjecture on the exact location
of the multicritical point in the phase diagram.

\subsection{General structure and conjecture}

In the homogeneous variables $x_k$,
the duality relation we have derived for the $\pm J$ Ising model is
(in the homogeneous partition function)
\begin{equation}
  Z_n(x_0, x_1, \cdots , x_n)=Z_n (x_0^{*}, x_1^{*},\cdots ,x_n^{*}).
  \label{dualZn}
\end{equation}
The question we try to answer in the present section is whether or not
it is possible to identify the transition point with a fixed point of duality transformation
for the replicated $\pm J$ Ising model and its generalizations (the Gaussian Ising model and
random $Z_q$ model) as is the case for the non-random ferromagnetic Ising model and the
standard scalar Potts model on the square lattice.
Mathematically rigorous argument on this problem can be developed only in a few limited cases.
We nevertheless take a step further to present a conjecture that the
fixed-point condition $x_0=x_0^*$ of only one of the $n+1$ independent
variables
has the possibility to lead to the exact transition point if we restrict ourselves
to the one-dimensional subvariety with enhanced symmetry, the NL.
Numerical data support our conjecture.

Let us first consider the generic case of arbitrary values of $x_0, x_1, x_2,\cdots , x_n$,
not just on the NL, until otherwise stated.
It is convenient to extract the factor $x_0$,
the edge Boltzmann factor for all-parallel spin state, in front of the partition function
in equation (\ref{dualZn}):
\begin{equation}
  Z_n(x_0, x_1, x_2, \cdots , x_n)\equiv (x_0)^{N_B} \Zc_n (u_1, u_2, \cdots , u_n),
\end{equation}
where $N_B$ is the number of bonds on the lattice, and
the reduced variables are defined by $u_k=x_k/x_0$.
This representation implies that we measure the energy relative to the all-parallel (perfectly
ferromagnetic) state.
The duality relation (\ref{dualZn}) is rewritten in terms of $\Zc_n$ as
 \begin{equation}
  (x_0)^{N_B}\Zc_n(u_1, u_2, \cdots , u_n)=(x_0^{*})^{N_B}
   \Zc_n(u_1^{*}, u_2^{*}, \cdots , u_n^{*}),
   \label{dualZn2}
 \end{equation}
where $u_k^{*}=x_k^{*}/x_0^{*}$.

\begin{figure}[hbt]
  \begin{center}
  \includegraphics[width=.58\linewidth]{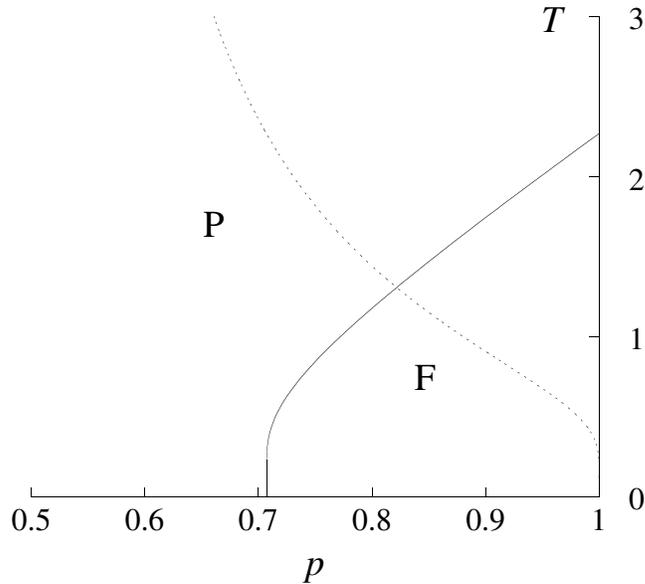}
  \end{center}
 \caption{Phase diagram of the $\pm J$ Ising model with $n=1$.
 The phase boundary is given by $x_0=x_0^*$ and the NL is shown dotted.}
 \label{fig:phase_diagram_n=1}
\end{figure}

When $n=1$, this form of duality allows us to identify the fixed point
with the transition point because the partition function has only one variable $\Zc (u_1)$,
and $u_1^*$ is a monotone decreasing function of $u_1$:
The transition point, if unique, should be at the fixed point $u_1=u_1^{*}$.
It then follows from equation (\ref{dualZn2}) that $x_0$ also satisfies the fixed point
condition $x_0=x_0^*$ because the functional values of $\Zc_n$ on both sides are equal
when the arguments $u_1$ and $u_1^{*}$ coincide.
We can therefore use the fixed point condition $x_0=x_0^*$ in place of $u_1=u_1^*$ to locate the transition point.
If we apply this argument to the $\pm J$ Ising model (the annealed model),
the condition $x_0=x_0^*$ defines a curve which is the exact phase boundary in the $p$-$T$
phase diagram as depicted in figure \ref{fig:phase_diagram_n=1}.

\begin{figure}[hbt]
  \begin{center}
  \includegraphics[width=.58\linewidth]{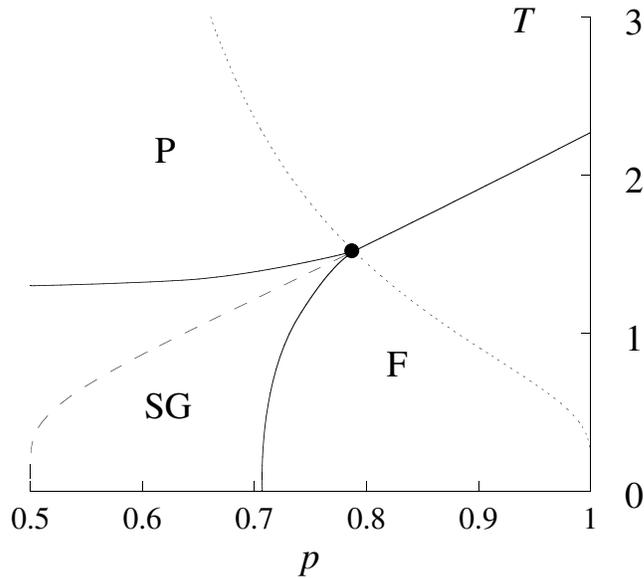}
  \end{center}
 \caption{Phase diagram of the $\pm J$ Ising model with $n=2$.
 The condition $x_0=x_0^*$ gives the correct phase boundary (full curve) above
 the multicritical point but not below (shown dashed).}
 \label{fig:phase_diagram_n=2}
\end{figure}

For $n=2$, the situation is not very different.
The condition $x_0=x_0^*$ is equivalent to $x_1=x_1^*$ as well as to $x_2=x_2^*$
as can be verified from equation (\ref{dual_2}).
Thus we have $u_1=u_1^*$ and $u_2=u_2^*$ as soon as we impose the fixed point
condition of the principal edge Boltzmann factor, $x_0=x_0^*$.
This fact is of course consistent with equation 
(\ref{dualZn2}) because both the prefactor
$(x_0)^{N_B}$ and the functional value $\Zc_n(u_1, u_2)$
 are the same on both sides
when $x_0=x_0^*$.
In the case of the $\pm J$ Ising model,
this fixed point condition $x_0=x_0^*$ defines a curve
 in the phase diagram drawn in
figure \ref{fig:phase_diagram_n=2}
(full curve above the multicritical point and dashed curve terminating at $p=0.5$).
As discussed by Georges \etal \cite{Georges}, this curve coincides with the
exact phase boundary above the multicritical point on the NL.
Below the multicritical point the phase boundary splits into two parts, one
separating the paramagnetic and spin glass phases and the 
other separating the ferromagnetic
and spin glass phases.
The duality transformation $x_0 \leftrightarrow x_0^*$ exchanges these
two phase boundaries; the uniqueness assumption of the transition point is
not satisfied in this case below the multicritical point.
The NL has a high symmetry since it satisfies $x_1=x_2$ as discussed
in equation (\ref{N_subvariety}), which is a globally self-dual subvariety because $n$ is even,
see equation (\ref{N_dual2}).
This means that the NL is mapped to itself by the duality transformation
and we are allowed to identity the transition point on the NL
(the multicritical point) with the fixed point, $x_0=x_0^*$, which indeed gives the
exact location of the multicritical point.
The lesson is that the condition $x_0=x_0^*$ does not always give the exact phase boundary
but it may make sense to search for the multicritical point on the NL by investigating
the intersection of the NL and the $x_0=x_0^*$ curve.

Another straightforward case is the general $2^n$-state standard scalar
Potts model which has $u_1=u_2=u_3=\cdots =u_n\equiv u$:
The reduced partition function  has then only a single variable $\Zc_n(u)$,
and we can identify the duality fixed point $u=u^*$ with 
the unique transition point.
It is worth pointing out that the condition $x_0=x_0^*$ together with equation (\ref{dualZn2})
automatically means that the values of $\Zc_n$ on both
 sides are equal $\Zc_n(u)=\Zc_n(u^*)$
and consequently the arguments are equal $u=u^*$.
Therefore the fixed point condition of the principal edge
 Boltzmann factor $x_0=x_0^*$ suffices
to identify the transition point (assuming it is unique) in 
the highly symmetric case
of the standard scalar Potts model.
Note that, when $n=2$, the $\pm J$ Ising model on the NL coincides
with the four-state standard scalar Potts model.

Let us next move on to the case of general $n$.
The subvariety $\cal N$ defined as the set of points satisfying
$u_1=u_{n}, u_2=u_{n-1}, u_3=u_{n-2},\cdots$ is transformed 
into the dual ${\cal N}^*$
satisfying $u_1^{*}=u_2^{*}, u_3^{*}=u_4^{*},\cdots $ as 
discussed in section \ref{sec:subvariety}.
In addition, the subvariety $\cal N$ is globally self-dual
${\cal N}={\cal N}^*$ ($u_1^*=u_{n}^*, u_2^*=u_{n-1}^*,\cdots$)
 when $n$ is even.
The number of independent variables of $\Zc_n$ is $[(n+1)/2]$ 
in $\cal N$, and it is not
possible to straightforwardly apply the argument developed
 above for the simple case of a single variable.
If we further restrict ourselves to the NL of the $\pm J$ Ising model,
we consider the one-dimensional curve in $\cal N$ representing the
the condition (\ref{N_cond}), which does not necessarily coincide 
with its dual
(except in the simplest cases of $n=1$ and 2).
Therefore it is difficult to use duality arguments to locate
 the transition point
for generic $n$ even if we restrict ourselves to the subvariety $\cal N$
(which satisfies a certain amount of symmetry) or its one-dimensional 
submanifold NL.

We nevertheless proceed by trying an ansatz that the fixed point condition
of the principal component of the edge Boltzmann matrix
 $x_0=x_0^{*}$ may lead to
the exact location of the transition point if we restrict
 ourselves to the NL
in the subvariety $\cal N$.
This ansatz is partly motivated by the above-mentioned example 
of the standard scalar Potts model,
in which the condition $x_0=x_0^*$ was sufficient to
 find the exact transition point.

One may wish to try a different fixed point condition 
like $x_1=x_1^*$ to investigate
the possible location of the transition point.
A physical reason to choose $x_0=x_0^*$, not the other
 ones, is that this edge Boltzmann factor $x_0$
is special in the sense that it is for the all-parallel  spin configuration,
and the expression of duality in equation (\ref{dualZn2}), 
which singles out
$x_0$ as a special variable, means to measure the energy
 relative to the perfectly ferromagnetic state.
This is natural when we discuss the properties of the system on the NL
which starts from the perfectly ferromagnetic ground 
state ($p=1, T=0$) and terminates
in the perfectly random high-temperature limit ($p=\frac{1}{2}, T\to \infty$) and
the multicritical point on this line separates the 
ferromagnetic and paramagnetic phases
\cite{Nishimori01}.
Another reason is that, if we try another condition,
 $x_1=x_1^*$ for example, the
result does not satisfy the rigorous inequality (\ref{Tcineq}).
This and other related points will be discussed in more detail later.

The present ansatz $x_0=x_0^{*}$ to locate the 
multicritical point on the NL is
equivalent to the one given in \cite{Nishimori-Nemoto}.

\begin{figure}[hbt]
  \begin{center}
  \includegraphics[width=.58\linewidth]{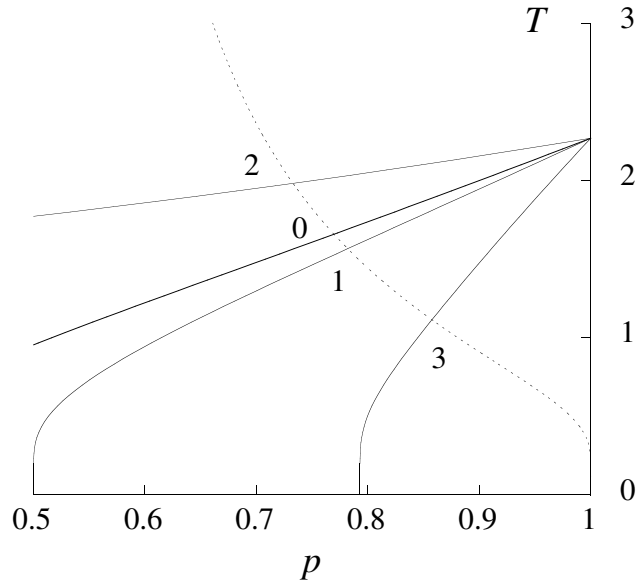}
  \end{center}
 \caption{The curves representing
 $x_0=x_0^{*}, x_1=x_1^{*}, x_2=x_2^{*}$ and $x_3=x_3^{*}$ marked 0, 1, 2, 3, respectively,
 for the $n=3$ model.}
 \label{fig:n=3}
\end{figure}

The condition $x_0=x_0^{*}$ is written explicitly for
 the $\pm J$ Ising model as,
using equations (\ref{generalxn}) and (\ref{xmstar_pmJ}),
 \begin{equation}
   p e^{nK} +(1-p) e^{-nK} =2^{-n/2} (e^{K} +e^{-K})^n.
    \label{fixed}
 \end{equation}
This relation, in conjunction with the NL condition (\ref{N_cond}), yields
 \begin{equation}
   e^{(n+1)K}+e^{-(n+1)K} =2^{-n/2} (e^K +e^{-K})^{n+1}.
    \label{fixed-K}
 \end{equation}
This equation, derived from $x_0=x_0^*$ plus the NL condition, gives the exact transition point (multicritical point) in the
cases of $n=1$ and 2 as already discussed.
When $n=3$, the curve given by equation (\ref{fixed}) is
drawn in figure \ref{fig:n=3} (marked 0) together with the curves coming
from three other conditions $x_1=x_1^{*}$, $x_2=x_2^{*}$ and $x_3=x_3^{*}$ marked 1, 2 and 3, respectively.
Apparently the intersections of the NL (dotted curve) and the four curves
corresponding to $x_0=x_0^*, x_1=x_1^{*}, x_2=x_2^{*}, x_3=x_3^{*}$
can be candidates of the transition point.
To check these possibilities, let us note that
equation (\ref{fixed-K}) coming from $x_0=x_0^*$ gives 
$T_{\rm c}^{(3)}=1.65858$
as a candidate of the multicritical point for $n=3$.
Similarly, $T_{\rm c}^{(4)}=1.75717$ for $n=4$, and
 $T_{\rm c}^{(5)}=1.82955$ for $n=5$
if we consistently use equation (\ref{fixed-K}).
When  $n=1$ and $n=2$, the same equation leads to $T_{\rm c}^{(1)}=1.30841$ and 
$T_{\rm c}^{(2)}=1.51865$, both
of which agree with the exact solution as they should.
These conjectured values of $T_{\rm c}^{(n)}$ from the 
condition $x_0=x_0^*$ plus the NL
all satisfy the rigorous inequality (\ref{Tcineq}):
 \begin{equation}
 T_{\rm c}^{(1)}<T_{\rm c}^{(2)}<T_{\rm c}^{(3)}
 <T_{\rm c}^{(4)}<T_{\rm c}^{(5)}.
 \end{equation}
If we instead assume that $x_1=x_1^{*}$ together with the NL 
condition (\ref{N_cond})
may give the multicritical point, the same inequality is violated at $n=4$
because, then, $T_{\rm c}^{(3)}=1.56655$ and $T_{\rm c}^{(4)}=1.51865$.
The same is true for the other $x_k=x_k^*~(k\ge 2)$.
Thus $x_0=x_0^*$ is the only possibility consistent with the inequality
if we are to choose the multicritical point among the intersections of $x_k=x_k^*~(k=0,1,\cdots ,n)$ and the NL.

It should be of some interest to discuss the limit
 $n\to \infty$ because the
problem can be solved exactly.
The $n$-replicated partition function for arbitrary $n$ is
 \begin{equation}
  Z_n =[Z(K)^n]_{\rm av} = [e^{-n\beta N f(K)}]_{\rm av}.
    \label{Znlarge}
 \end{equation}
If we take the limit $n\to\infty$ with $N$ (the number
 of sites) kept large but finite,
the number of terms in the above configurational average remains finite, and
the value of this equation is dominated by the term with 
the smallest value of $f(K)$.
The smallest value of the free energy is expected to be
 given by the perfectly ferromagnetic
bond configuration (and its gauge equivalents), and therefore we have
 \begin{equation}
   Z_n \approx e^{-\beta_p N f_0(K_p)}e^{-n\beta Nf_0(K)},
 \end{equation}
where $f_0$ is the free energy of the non-random ferromagnetic Ising model.
The first factor on the right-hand side is the probability weight of the
non-random configuration and its gauge equivalents (proportional to $Z(K_p)$;
see section \ref{sec:gauge}).
The system then has a unique critical point at the 
non-random critical point $K=K_{\rm c}^{\rm F}$
as well as at $K_p=K_{\rm c}^{\rm F}$ (the latter being
 equivalent to $p_{\rm c}=\frac{\sqrt{2}}{2}$).
The phase boundaries are two lines representing these two critical parameters.
On the other hand, the condition $x_0=x_0^{*}$ gives in 
the limit $n\to \infty$
 \begin{equation}
   e^{-2K}=\sqrt{2}-1
 \end{equation}
which agrees with the above analysis giving $K=K_{\rm c}^{\rm F}$.
Thus the conjecture that the multicritical point is given by the intersection
of $x_0=x_0^*$ and the NL is correct in this limit $n\to\infty$.
It is interesting to note that for $n\to\infty$
all the conditions $x_k=x_k^{*}$ with $k$ finite lead to the same conclusion.
However, for finite $n$, the result depends on $k$, and the $k=0$ condition is
the only one consistent with the inequality (\ref{Tcineq}).

\subsection{Numerical evidence for $n=3$}

We have carried out an extensive Monte Carlo simulation of the $n=3$ $\pm J$ Ising model
on the NL on the square lattice to check the above-mentioned  possible value of the
multicritical point $T_{\rm c}^{(3)}=1.65858$.
A preliminary scan of the energy showed that the transition is very likely to be of
first order since clear hysteresis has been found.
This is natural because the system has eight($=2^3$) degrees of freedom at
each site and is not very far from the eight-state standard scalar Potts model
(which has a first-order transition)
because of the enhanced symmetry $x_1=x_3$ on the NL
(a little short of the high symmetry $x_1=x_2=x_3$ of the standard scalar Potts model).

\begin{figure}[hbt]
  \begin{center}
 \includegraphics[width=.70\linewidth]{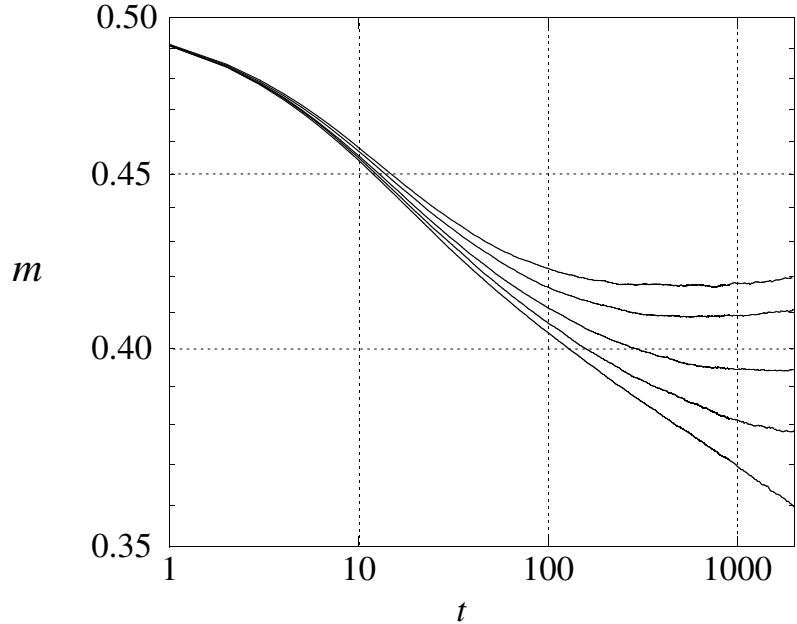} 
  \end{center}
 \caption{Non-equilibrium relaxation of magnetization with
 mixed phase initialization for the $n=3$ $\pm J$ Ising model on the NL on
 the square lattice. The temperatures
 are 1.650, 1.655, 1.660, 1.663, 1.665 from
 top to bottom.}
 \label{fig:NER}
\end{figure}

We have therefore used the non-equilibrium relaxation method (\cite{Ozeki,Ito}
and references therein),
which allows us to simulate very large systems in its initial relaxation
stage to investigate equilibrium properties.
Another reason to use the non-equilibrium relaxation method is that it
allows us to identify the first-order transition point without
evaluating the free energy \cite{Ozeki}.
The results for the relaxation of magnetization
are drawn in figure \ref{fig:NER} for the temperature range
from $T=1.650$ (top curve) to $T=1.665$ (bottom curve)
with linear system size $L=8000$, averaged over several samples,
under mixed phase initialization appropriate for first-order
transition \cite{Ozeki}.
We have confirmed by comparison with smaller systems (and larger systems ($L=12000$) in
some limited cases)
that this size $L=8000$ is sufficiently large to investigate
the properties of infinite-size systems at least to the time steps
indicated in figure \ref{fig:NER} (2000 steps).
From the positive slope and upward curvature of magnetization after about 1000 Monte Carlo steps for $T=1.655$,
we conclude that this temperature is in the ferromagnetic phase
due to the prescription of the non-equilibrium relation method \cite{Ozeki}.
Similarly the system is judged to lie in the paramagnetic phase at $T=1.665$ from
the negative slope and downward curvature.
Our conclusion is $T_{\rm c}^{(3)}=1.660(5)$, which is consistent with the
above-mentioned conjecture 1.65858.
The values from other fixed point conditions are completely ruled out,
1.56655 (from $x_1=x_1^*$), 1.98207 ($x_2=x_2^*$) and 1.11466 ($x_3=x_3^*$).

\subsection{Quenched limit}
There have been a number of numerical studies of quenched systems ($n \to 0$)
to find the location of multicritical point on the square lattice.
If we take the limit  $n \to 0$ in equation (\ref{fixed-K}), 
the formula rewritten in terms of $p$ acquires a simple and appealing expression
suggesting that exactly half of the entropy of bond distribution is exhausted
at $p_{\rm c}$:
 \begin{equation}
  -p_{\rm c} \ln p_{\rm c} -(1-p_{\rm c})\ln (1-p_{\rm c} )=\frac{\ln 2}{2}.
  \label{pc}
 \end{equation}
Numerically this equation gives $p_{\rm c}=0.88997$, which is compared very favourably
with numerical results, among which 0.8894(9) is the most recent and extensive one \cite{Ito}, as
well as 0.8905(5) \cite{AaraoReis}, 0.886(3) \cite{Singh},
0.8906(2) \cite{Honecker} and 0.8907(2) \cite{Merz}.

\begin{figure}[hbt]
  \begin{center}
 \includegraphics[width=.60\linewidth]{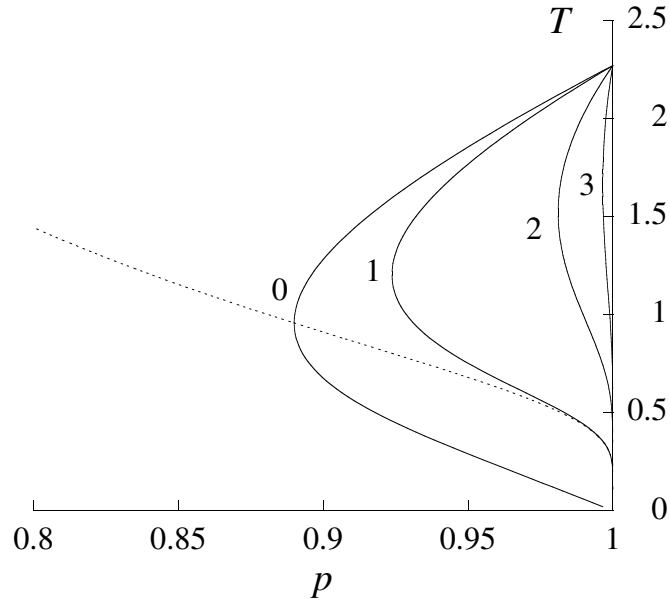} 
  \end{center}
 \caption{The curves representing fixed point conditions $x_0=x_0^*$,
 $x_1=x_1^*$, $x_2=x_2^*$ and $x_3=x_3^*$ marked 0, 1, 2 and 3, respectively,
 in the quenched limit.  The NL is shown dotted.}
 \label{fig:quenched}
\end{figure}

Interestingly, the curve (\ref{fixed}) coming from $x_0=x_0^*$ is the only one with
an intersection with the NL in the quenched limit.  All the other curves
$x_k=x_k^*~(k\ge 1)$ do not cross the NL as depicted in figure \ref{fig:quenched}.

If we further apply the condition $x_0=x_0^{*}$ to a model with continuous
distribution of coupling as in equation (\ref{xk}), we find in the limit $n\to 0$
 \begin{equation}
   \int du\, P(u)\, \ln (1+e^{-2\beta u})=\frac{\ln 2}{2}.
   \label{Ising_boundary}
 \end{equation}
This expression already appeared in \cite{Nishimori79}.
Under the NL condition (\ref{generaln}), this equation reads
 \begin{equation}
   \int du e^{\beta u}F(u^2) \ln (1+e^{- \beta u})=\frac{\ln 2}{2},
 \end{equation}
which gives for the Gaussian model $J_0/J=1.02177$.
Numerical results are consistent with this value, 1.00(2) (Y Ozeki \etal, private communication).

For the random chiral Potts model with
 \begin{eqnarray}
   &&V(0)=K,~V(1)=V(2)=\cdots =V(q-1)=0\\
   && p_0=1-(q-1)p, ~p_1=p_2=\cdots p_{q-1}=p
   \label{chiralPotts}
 \end{eqnarray}
under the NL condition \cite{Nishimori83}
 \begin{equation}
   p=\frac{1}{e^{K}+q-1},
 \end{equation}
the conjectured multicritical point coming from $x_0=x_0^*$ is in the quenched limit,
using equations (\ref{xkq}) and (\ref{xkqd}),
 \begin{equation}
   -(1-(q-1)p_c) \ln (1-(q-1)p_c)-(q-1)p_c \ln p_c =\frac{\ln q}{2}.
   \label{Pottspc}
 \end{equation}
Note that the model with equation (\ref{chiralPotts}) reduces to the non-random case
in the limit $p\to 0$ in contrast to our previous convention of $p\to 1$;
the reason is to avoid confusion when we compare our conjecture
with the numerical investigation which suggests $p_{\rm c}$ between 0.079
and 0.080 for $q=3$ \cite{Jacobsen}.
Equation (\ref{Pottspc}) yields $p_{\rm c}=0.079731$ for $q=3$.
We therefore conclude that the condition $x_0=x_0^*$ has been confirmed to give the correct
transition point on the NL within numerical accuracies for all available cases.

\section{Summary and Discussions}

We have derived symmetries of the models of the two-dimensional spin glass.
In particular, invariances of the replicated partition function $Z_n$ under
 exchanges of edge Boltzmann factors and
duality have been proved.
Also discussed is the invariance of the structure of the edge Boltzmann matrix by inversions.
These properties hold for generic values of edge Boltzmann factors,
not just for the replicated $\pm J$ Ising model.
The dual Boltzmann factor of the $\pm J$ Ising model
has an interesting expression in terms of dual spin variables.
Griffiths inequalities apply to this expression, leading to monotonicity of phase boundaries,
equivalent to monotonicity of the location of multicritical point as a function of $n$.

Complexities under inversions have been investigated and a remarkable
reduction of complexities (enhanced symmetry) has been observed in the subvariety $\cal N$.
This result suggests that the behaviour of the system is simpler in this subvariety
than at generic points even if the problem is not (generically) integrable
because the exponential growth excludes integrability.

Conjecture on the exact location of the multicritical point on the NL has been presented
based on the duality and symmetry arguments.
Reasons have been explained that the intersection of the curve $x_0=x_0^*$ and
the NL is the only plausible candidate of the multicritical point among a set of
similar conjectures ($x_k=x_k^*$ plus the NL).
Numerical results
are in very good agreement with this conjecture.
Nevertheless, recalling the situation already encountered for the
analysis of the phase diagram of the isotropic 
three-state chiral Potts model~\cite{MaWuHu92a,MaWuHu92}, where 
the critical manifold is numerically extremely close to the
 self-dual condition $\, x_0^{*} \, = \, x_0$
but is possibly mathematically different (in contrast with the situation
encountered in the symmetric Ashkin-Teller model), one cannot 
discard the possibility that, restricted to the NL,
 the transition point could be 
numerically extremely close to the intersection with 
 $\, x_0^{*} \, = \, x_0$ but, actually, mathematically different.
These points need to be further investigated carefully.


It may be useful to consider the possibility that the condition $x_0=x_0^*$ could
be a very good approximation of the transition manifold away from the intersection
with $\cal N$ but in the high temperature part of the phase boundary as in figure
\ref{fig:phase_diagram_n=2} at least for random ferromagnetic spin systems.
For the $q$-state random standard ferromagnetic Potts model (the random version of the
ferromagnetic scalar Potts model), the condition $x_0=x_0^*$ in the quenched limit is
\begin{eqnarray}
\label{boundary}
\int d\mu(e^{K}) \, \ln\Bigl({{e^{K}\, + (q-1)} \over {e^{K}}}\Big) 
\,\, \, \,\,    = \, \,\,  \, \,\,    {{1} \over {2}} \ln q,
\end{eqnarray}
where the positive interaction $K$ is assumed to be distributed with measure $d\mu(e^{K})$.
This is a Potts-generalization of equation (\ref{Ising_boundary}).
Investigation of the consequences of this equation is going on.

Another interesting observation is that the formula (\ref{fixed-K}) gives $K$ as a function
of $n$ which diverges as $n$ approaches $-1$.
It is not obvious at all that we can apply this formula to such a limit, but {\em if} we do so,
then the divergence of $K$ implies that the transition point on the NL is
$T_{\rm c}^{(n)}\to 0$ as $n\to -1$.
If we use the relation between $n$ and $n+1$ in equation (\ref{ZnZn1}), we find
that the transition point of the $n\to 0$ (quenched) system vanishes at $p=\frac{1}{2}$.
Although we should be very careful to apply our results to such a limit of negative $n$,
this last result is reasonable and interesting in its own right
because it supports the usual consensus that the two-dimensional Ising spin glass
does not have a finite-temperature phase transition
when the distribution of bond randomness is symmetric.
More efforts should be devoted to the investigation of this problem.

Let us also comment here on the equivalence between the $n$-replicated 3$d$ random lattice gauge theory
and its dual, the $n$-replicated Ising spin ferromagnet with edge Boltzmann factor
described in section \ref{sec:DualBF}.  All the arguments  sections \ref{sec:DualBF} and 
\ref{sec:IneqBF} apply to such a case.
In particular the bound on the critical probability like (\ref{pc-bound2}) results:
$p_{\rm c}^{(0)}({\rm MCP})\ge 0.9005$.  Numerically it is 0.97 \cite{Wang} \footnote{
This value is for the $T=0$ transition point, not for the multicritical point,
but is likely to be close to the latter.
}.

We have seen that the
 replica analysis naturally yields to consider a class 
of remarkable (non-random) lattice spin models, defined by equation (\ref{general_BM}), 
which are highly structured
and for which extensive symmetry analysis can be performed
exactly. Recalling Domany's $\, (N_{\alpha}, \,N_{\beta})\, $
 terminology \cite{Domany}, these models correspond to highly symmetric 
specific $\, (N_{2}, \cdots , \,N_{2})$ models
and, more generally $\, (N_{q}, \cdots , \,N_{q})$ models.
These singled-out classes of lattice spin edge models 
provide a very powerful tool of analysis for 
many spin-glass problems and are also worth studying {\em per se}. 

One should note that many of the exact calculations, 
displayed in this paper, are still valid 
when the distribution of the coupling constants
is not of  the $\, \pm J$ type
or some continuous distribution like equation (\ref{generaln})
but, for instance, a two-delta-peaks $\, (J_1, \, J_2)$ distribution,
or, even, a totally general distribution : We only need to 
get an effective Boltzmann matrix of the hierarchical
form  (\ref{general_BM}).
One should also underline that these models can  
straightforwardly be generalized to chiral Potts models (see \ref{sec:AppendixB}),
and also to spin models without any Wu-Wang duality (like the Ising
 model with a magnetic field, see \ref{sec:AppendixB} ; the non-linear
inversion relations $\, I$ and $\, J$ taking the place of the linear
duality transformation $\,D$), providing room for many new exact results
on spin-glass problems. 

\ack
The work of HN was supported by the Grant-in-Aid for Scientific Research
by the Ministry of Education.

\appendix
\setcounter{section}{0}
\section{}\label{sec:AppendixA}
In this appendix we outline the proof of equation (\ref{N_dual2}) assuming equation (\ref{N_subvariety})
and the symmetry (\ref{Zn_mirror}) under the operation $M$.
According to the duality relation (\ref{x_star}) for even $n=2q$, the expression
of $x_m^*$, after applying the operation $M$, reads
 \begin{eqnarray}
   2^q \, x_m^*&=&\sum_{k=0}^q D_{m}^{2k}x_{2k}+\sum_{k=1}^q D_{m}^{2k-1}x_{2k-1}
   \nonumber\\
  & \rightarrow &
   \sum_{k=0}^q D_{m}^{2k}x_{2k}+\sum_{k=1}^q D_{m}^{2k-1}x_{2q-2k+1}\nonumber\\
  &=&D_m^0x_0 +D_m^2 x_2 +D_m^4 x_4+\cdots +D_m^{2q}x_{2q} \nonumber\\
  && +D_m^1 x_{2q-1}+D_m^3 x_{2q-3}+\cdots +D_m^{2q-1} x_1.
  \end{eqnarray}
We then impose the condition (\ref{N_subvariety}) to find
 \begin{equation}
   2^q\, x_m^*=D_m^0 x_0 +(D_m^1+D_m^2)x_2 +(D_m^3+D_m^4)x_4 +\cdots +(D_m^{2q-1}+D_m^{2q})x_{2q}.
 \end{equation}
Thus, in order to show $x_m^*=x_{2q-m+1}^*$, it suffices to derive
 \begin{equation}
   D_{m}^{2k-1}(2q)+D_m^{2k}(2q)=D_{2q-m+1}^{2k-1}(2q)+D_{2q-m+1}^{2k}(2q),
   \label{Dm1}
 \end{equation}
where we have written the $n(=2q)$-dependence explicitly.
This equation can be proved by induction with respect to $q$.

The following relation will be useful for the proof:
 \begin{equation}
  D_{m+1}^k(n)+D_{m+1}^{k-1}(n)=D_m^k (n)-D_m^{k-1}(n)
 \label{Dm_red2}
 \end{equation}
which is derived by replacing $m$ in equation (\ref{t-expansion}) with $m+1$
(which amounts to multiplying both sides by $(1-t)/(1+t)$ ).

It is easy to check explicitly using equation (\ref{Dmk})
that the target relation (\ref{Dm1}) is valid for small $q$
with any $m$ and $k$.
Let us then assume that this equation holds for $q$ with any $m$ and $k$ and show that the same is true for $q+1$.
The left-hand side of equation (\ref{Dm1}) with $2q$ replaced by $2q+2$ can be
reduced to an expression with $2q$ using the recursion relation (\ref{D-recursion1}) twice,
 \begin{eqnarray}
  && D_m^{2k-1}(2q+2)+D_m^{2k}(2q+2)  \nonumber\\
  &=&D_m^{2k}(2q)+D_m^{2k-1}(2q)
     +2 (D_m^{2k-1}(2q)+D_m^{2k-2}(2q)) \nonumber\\
    &&  +D_m^{2k-2}(2q)+D_m^{2k-3}(2q).
 \end{eqnarray}
Application of the other recursion relation (\ref{Dm_red2}) to each pair of terms
on the right-hand side of the above equation yields
 \begin{eqnarray}
   &&D_m^{2k-1}(2q+2)+D_m^{2k}(2q+2)\nonumber\\
   &=& D_{m-1}^{2k}(2q)+D_{m-1}^{2k-1}(2q)-D_{m-1}^{2k-2}(2q)-D_{m-1}^{2k-3}(2q).
  \label{lhs}
 \end{eqnarray}
This is our expression for the left-hand side of equation (\ref{Dm1}) with $q\to q+1$.
The right-hand side of equation (\ref{Dm1}) with $2q$ replaced by $2q+2$ can also be rewritten
by the recursions (\ref{D-recursion1}) and (\ref{Dm_red2}) to reach a similar expression
 \begin{eqnarray}
  &&D_{2q+2-m+1}^{2k-1}(2q+2)+D_{2q+2-m+1}^{2k}(2q+2)\nonumber\\
   &=& D_{2q-m+2}^{2k}(2q)+D_{2q-m+2}^{2k-1}(2q)-D_{2q-m+2}^{2k-2}(2q)-D_{2q-m+2}^{2k-3}(2q).
 \end{eqnarray}
This equation is equal to equation (\ref{lhs}) by the starting assumption of induction,
which completes the proof.

\section{}\label{sec:AppendixB}

The results on the Ising model generalize straightforwardly to $\, q$-state models.
Let us consider, for instance, 
 the three-state chiral 
Potts model corresponding to a $\, 3 \times 3$ cyclic 
edge Boltzmann matrix\footnote{That is the  chiral 
Potts model studied by Baxter, Perk and
 Au-Yang~\cite{BaPeAu88}.}. For an arbitrary
 number $\, n$ of replicas,
 the previous hierarchical scheme generalizes
 straightforwardly. If one denotes by $\, A_n\, $
the  $\, q^n \times
 q^n\, $ effective edge Boltzmann
 matrices corresponding to $\, [ Z^n ]_{\rm av}\, $
they can be obtained by the following recursion :
\begin{eqnarray}
\label{hierarch}
A_n \, \,  \rightarrow \quad 
A_{n+1}\, \, = \, \, 
\left[ \begin {array}{ccc} 
A_n&B_n&C_n\\
\noalign{\medskip}C_n&A_n&B_n\\
\noalign{\medskip}B_n&C_n&A_n
\end {array} \right],  
\end{eqnarray}
where 
\begin{eqnarray}
A_1 \, = \, \left[ \begin {array}{ccc} 
x_{0,0}&x_{0,1}&x_{1,0}\\
\noalign{\medskip}x_{1,0}&x_{0,0}&x_{0,1}\\
\noalign{\medskip}x_{0,1}&x_{1,0}&x_{0,0}
\end {array} \right] \quad \quad \quad \quad 
\end{eqnarray}
and
\begin{eqnarray}
B_n \, = \, \, A_n(x_{m,p} \, \rightarrow  \,x_{m,p+1}) , \qquad
C_n \, = \, \, A_n(x_{m,p} \, \rightarrow  \,x_{m+1,p}) . \qquad\nonumber 
\end{eqnarray}
 Note that the number 
of homogeneous parameters necessary
to describe  the pattern (\ref{hierarch})
grows {\em quadratically} with the number
 of replicas like $\, (n+2)\cdot (n+1)/2$
(this number is $3$ for $\, n\, = \, 1$, 
$6$ for $\, n\, = \, 2$,
$10$ for $\, n\, = \, 3$, $15$ for $\, n\, = \, 4$,
$21$  for $\, n\, = \, 5$, ...). The family of
 $\, 3^n \times 3^n$ models we have to study is thus
a little more complicated (as well as the equivalent of 
the subvarieties ${\cal N}$ (\ref{N_subvariety})). 
This is in contrast with the $\, q\, = \, 2$ Ising model without
magnetic field,
analysed in the text, for which the number of homogeneous parameters
$\, x_k\, $ grows linearly with $\, n$ (this number is $\, n+1$).  
More generally, 
for a cyclic $\, q \times q\, $ matrix, the number of 
homogeneous parameters necessary
to describe a pattern generalizing (\ref{hierarch})
will be the number of partition of $\, n$
into $\, q$ natural integers $\, k_i$ :
$\, k_1\, = \, k_2 \, + \, k_3 \, + \cdots \, + k_q \, = \, n$
with $\, k_i \, = \, 0, \, 1, \, \cdots n$.
The duality transformation on these models (\ref{hierarch}) is 
inherited from the duality transformation on the three-state chiral
Potts model associated with cyclic
$\, 3 \times 3 $ matrices.

Let us consider a simple example of spin edge
model with no duality transformation, namely the Ising model with a
magnetic field. The effective Boltzmann matrices can be obtained 
similarly by a hierarchical scheme generalizing (\ref{general_BM}) :
\begin{eqnarray}
\label{hierarch2}
A_n \, \,  \rightarrow \quad 
A_{n+1}\, \, = \, \, 
\left[ \begin {array}{cc} 
A_n&B_n\\
\noalign{\medskip}B_n&C_n
\end {array} \right], 
\end{eqnarray}
where
\begin{eqnarray} 
A_1 \, = \, \left[ \begin {array}{ccc} 
x_{0,0}&x_{1,0}\\
\noalign{\medskip}x_{1,0}&x_{0,1}
\end {array} \right] \quad \quad \quad \quad 
\end{eqnarray}
and
\begin{eqnarray}
B_n \, = \, \, A_n(x_{m,p} \, \rightarrow  \,x_{m+1,p}) , \qquad
C_n \, = \, \, A_n(x_{m,p} \, \rightarrow  \,x_{m,p+1}) . \qquad\nonumber 
\end{eqnarray}
Again the number 
of homogeneous parameters necessary
to describe  the pattern (\ref{hierarch2})
 grows  quadratically with the number
 of replicas like $\, (n+2)\cdot (n+1)/2$.
In this last case,  {\em we do not have a 
duality transformation $\,D$ anymore} : The
inversion relations $\, I$ and $\, J$ take
 the place of the linear
duality transformation $\,D$. However, the inverse
of matrices given by (\ref{hierarch2}) are not matrices 
of the same form (\ref{hierarch2}). One thus needs 
to consider slightly more general matrices than (\ref{hierarch2})
in order to be able to use the non-linear symmetry $\, I$.
Details will be given elsewhere.


\end{document}